%% file: template.tex
\documentclass{article}

\usepackage{arxiv}

\usepackage{graphicx}
\usepackage{amsmath,amsfonts,amssymb}
\usepackage{array}
\usepackage{textcomp}
\usepackage{stfloats}
\usepackage{url}
\usepackage{hyperref}
\usepackage{verbatim}
\usepackage{cite}
\usepackage{comment}
\usepackage{tablefootnote}
\usepackage[export]{adjustbox}
\usepackage{siunitx}
\usepackage{booktabs}
\usepackage{varwidth}
\usepackage[table]{xcolor}
\usepackage{bbding}
\usepackage{threeparttable}

\title{(In)Security of Mobile Apps in Developing Countries: A Systematic Literature Review}

\author{
 Alioune Diallo \\
  University of Luxembourg \\
  \texttt{alioune.diallo@uni.lu} \\
   \And
 Jordan Samhi \\
  School of Coumputing and Information\\
  CISPA Helmholtz Center for Information Security, Germany \\
  \texttt{jordan.samhi@cispa.de} \\
  \And
 Tegawendé Bissyandé \\
   University of Luxembourg \\
  \texttt{tegawende.bissyande@uni.lu} \\
  \And
 Jacques Klein \\
   University of Luxembourg \\
  \texttt{jacques.klein@uni.lu}
}

\begin{document}

\maketitle

\begin{abstract}
In developing countries, several key sectors, including education, finance, agriculture, and healthcare, mainly deliver their services via mobile app technology on handheld devices. 
As a result, mobile app security has emerged as a paramount issue in developing countries.
In this paper, we investigate the state of research on mobile app security, focusing on developing countries.
More specifically, we performed a systematic literature review exploring the research directions taken by existing works, the different security concerns addressed, and the techniques used by researchers to highlight or address app security issues. 
Our main findings are: 
(1) the literature  includes only a few studies on mobile app security in the context of developing countries;
(2) among the different security concerns that researchers study, vulnerability detection appears to be the leading research topic; and
(3) FinTech apps are revealed as the main target in the relevant literature. 
Overall, our work highlights that there is largely room for developing further specialized techniques addressing mobile app security in the context of developing countries.
\end{abstract}


\input{intro}
\input{background}

\input{methodology}
\input{result}
\input{discussion}
\input{threat_to_validity}
\input{related_work}
\input{conclusion}

\bibliographystyle{unsrt}  
\bibliography{references}  






\input{appendix}

\end{document}

%% file: intro.tex
\section{Introduction}
\label{intro}

Technological advancement has radically changed our daily lives in recent decades, providing tools such as tablets and smartphones.
Mobile use worldwide has seen rapid adoption in the last decade~\cite{olson2022smartphone, ash2023_bis}.
In particular, with 255 billion app downloads in the world~\cite{ash2023}, smartphones have become ubiquitous and used for various tasks. 

A developing country, also known as an underdeveloped, low-income, or middle-income country, is generally defined as a country that has not yet achieved full maturity in terms of economic and industrial development~\cite{DevpingCountries}.
The United Nations Development Programme (UNDP) utilizes the Human Development Index (HDI)\footnote{\url{https://hdr.undp.org/data-center/human-development-index\#/indicies/HDI}} score to assess whether a country falls into the category of developing or developed. 
In developing countries, mobile technology has enabled leapfrogging for mass adoption of critical digital services such as mobile money, e-health, e-agriculture, etc. Studies even forecast the adoption of smartphones in Sub-Saharan Africa to 87\% by 2030\footnote{Source: \url{http://www.osiris.sn/En-Afrique-subsaharienne-le-taux-d.html}}. 

While the popularity of smartphones worldwide has brought numerous opportunities and benefits \cite{keyideas2017}, it also comes with challenges and concerns due to the risks that they carry~\cite{rSheldon2019}. 
Mobile security threats and app vulnerabilities can pose significant risks in the critical services used.

On the one hand, mobile apps process sensitive data such as user credentials, financial data, medical information, user's location, etc.~\cite{alaiad2019determinants, pentina2016exploring}.
This makes users vulnerable to targeted attacks~\cite{pentina2016exploring}.
On the other hand, individuals are developing malicious apps that specifically target developing countries.
This is justified by the fact that most of the ten countries in 2022 mostly affected by mobile malware are developing countries~\cite{statica001}.
Additionally, mobile malware affects more than 10\% of developing country mobile devices~\cite{carter2017forces}.

Developing countries face unique challenges in many domains, particularly in the digital ecosystem~\cite{6859817}, which directly impacts mobile app security. These challenges include:
\begin{itemize}
    \item Limitations in connectivity, which hinder systematic software updates and lead developers to implement unsafe workarounds such as naive, local, and insecure authentication.
    \item Persaviness of low-cost, but insecure, devices, which are often infected with viruses right out of the box~\cite{kaspersky-web}.
    \item Low computing literacy and education, which creates awareness issues with security best practices in app development and usage.
\end{itemize}
These challenges contribute to an elevated level of digital security risks. 
Indeed, when developing mobile apps, developers may unintentionally leave security gaps that malicious actors can exploit, and outdated technology usage can compound security issues.
For example, in developing countries, mobile financial applications face numerous security challenges, including those related to encryption, authentication, management of keys, credentials, and other secrets~\cite{approov2023}.
There are also some frauds related to mobile financial services, such as theft of mobile money customer data using malware, account hijacking with SIM swap and changing MSISDN linked to the mobile money account, technical attacks such as DoS attack and transmission data interception via man in the middle targeting mobile money systems, and SMS Spoofing~\cite{fouad2017, Castle}.

Considering these challenges and the security aspects mentioned above, the security of mobile apps is a critical concern in developing countries.
Focusing on mobile app security in these contexts is vital for improving knowledge and enhancing defense against privacy violations, threats, and cyberattacks.

Researchers have performed various studies to tackle these security concerns: exploring the impact of requested permissions on the app vulnerability~\cite{Koala}, analyzing the source code to look for errors and misuses leading to potential vulnerabilities~\cite{Ibrar}, studying the app behaviors to identify those that are malicious~\cite{Bassole}, and looking for traces left by mobile apps in the device's memory after execution~\cite{Osho}.
Several secondary studies have been performed in the context of developing countries. 
Hoque et al.~\cite{hoque2020mobile} performed a review of studies related to mobile health applications by evaluating the quality of evidence reporting in mobile health literature, using the mobile health Evidence Reporting and Assessment (mERA) checklist as recommended by the World Health Organization (WHO). In this study, authors reveal a low level of familiarity with the mERA checklist among researchers in developing countries, and most mHealth studies do not adequately meet the essential criteria for evidence reporting.
Msweli et al.~\cite{msweli2020enablers} reviewed the existing literature about the adoption of mobile banking services among the elderly in developing countries, focusing on the enablers and barriers. The paper highlights that while mobile banking has become prevalent in both developed and developing nations, research focusing on its adoption among the elderly is limited. The main barriers identified include security concerns, trust and privacy issues, a lack of personalization, and limited technical knowledge among the elderly. On the other hand, significant enablers include the perceived ease of use of mobile banking applications, perceived value, convenience, and consumer attitudes. 
Malik~\cite{malik2020review} reviewed empirical research on Internet and mobile banking adoption in developing countries. The review focuses on the factors influencing the adoption of these technologies and the methodologies used in the studies. It also discusses the variables that have been extended in the Unified Theory of Acceptance and Use of Technology (UTAUT) model. In the study, the author highlights the directions, the most used analysis tools, and the main indicators of behavioral intention used in the publications.
There exist several other secondary studies, but none of them primarily focus on the security of mobile apps in developing countries considering all types of apps.
This means we still need a study providing a comprehensive overview of the current state of knowledge in this domain.
Furthermore, discerning unexplored research avenues is also a desirable task.

Other studies exist to investigate the security of mobile apps. For instance, Martinez-Perez et al.~\cite{martinez2015privacy} evaluate the current state of privacy and security in mobile health (mHealth) apps, focusing on reviewing existing laws regulating privacy and security in the European Union (EU) and the United States (USA), analyzing the corresponding academic literature, and proposing recommendations for app designers to ensure compliance with current security and privacy legislation. The review identified several research lines, including secure systems proposals, authentication techniques, and privacy aspects in Body Sensor Networks (BSNs). Key findings in this study highlighted the need for better security mechanisms in mHealth apps and the importance of user consent and data protection.
Other authors reviewed existing literature to identify and analyze the security issues associated with the use of mobile educational apps~\cite{mkpojiogu2021security}. The goal is to highlight the security challenges and propose ways to address these issues to enhance the security and usability of mobile educational apps. Their review identified several key security aspects in the use of mobile educational apps, including reliability, integrity, trust, privacy/confidentiality, and availability.
They suggested several ways to address these security challenges.
However, these papers do not focus on developing countries, and thus, they do not highlight the challenges specific to developing countries.

To fill these gaps in the context of developing countries, we conducted a systematic literature review (SLR)  by examining the existing works on mobile app security targeting mobile apps in developing countries.
Many studies have focused on security concerns in the specific context of developing countries.
For instance, since mobile money is based on USSD-embedded technology instead of apps, studies have explored SMS-based and SIM-based security issues~\cite{mambina2022classifying, 10.1145/3209811.3209817}.
However, few studies have focused on issues related to mobile applications.
In this study, we have excluded countries with HIGH and VERY HIGH HDI and we only focus on those considered to have MEDIUM and LOW HDI. This choice is based on the classification given by the UNDP in its Human Development Report\footnote{\url{https://hdr.undp.org/sites/default/files/2021-22_HDR/HDR21-22_Statistical_Annex_HDI_Table.xlsx}}.
Our focus on these countries is driven by our goal to provide actionable and impactful information for those tackling mobile application security issues in contexts where resources are scarcer, and challenges can be more pronounced.
Our work aims to identify the research directions and the literature gaps to highlight the directions that require further research.

The main contributions of this study are as follows:
\begin{itemize}
    \item We conduct a systematic search to map the research literature on mobile app security that targets developing countries.
            Through the systematic identification of existing studies, we offer an exhaustive survey of the field. This covers a broad range of pertinent research, thereby establishing a robust groundwork for future investigations.
    \item We review the identified literature and provide analysis on: 
    \\(1) the type of research performed (e.g., user study, app analysis, study of the development framework, app security testing, or study that focuses on proposing design \& implementation solution) - Reviewing the identified literature and analyzing the types of research performed is paramount. No study explicitly focuses on this aspect within the context of developing countries. Our analysis sheds light on the methodologies used, allowing researchers to understand the landscape better; 
    \\(2) the security concerns that have been addressed - Investigating the security concerns addressed in the literature is important. Understanding which vulnerabilities and threats are studied helps identify gaps and prioritize research efforts. This focus aims to identify the specific security concerns relevant to developing countries, which helps highlight their unique challenges. It contributes to the field by tailoring solutions to local contexts; and 
    \\(3) the type of analysis performed - Identifying and categorizing the techniques employed to assess mobile app security provides a valuable resource for researchers and practitioners. 
    \item We identify research directions that must be followed to contribute to delivering more secure mobile apps in developing countries.
            By suggesting specific areas for improvement, we guide future research efforts. 
            This encourages the development of specialized techniques that directly address the needs of these regions.
\end{itemize}

By combining these elements, we provide a good and nonexistent resource for researchers, practitioners, and policymakers aiming to enhance mobile app security in regions with unique challenges.

\textbf{Data Availability Statement.} The datasets used in the current study are available in our online repository~\cite{dialloDatasetSLR2024}.
\\

\textbf{Paper Organization.}
The rest of this paper is organized as follows. 
Section~\ref{backgrnd} presents the background. 
Section~\ref{method} introduces the detailed methodology followed in this SLR. 
Section~\ref{result} reports the findings. 
Section~\ref{discuss} discusses explored and unexplored research directions, publication trends, and future challenges, 
Section~\ref{validity} presents the threat that can compromise the validity of this study.
Section~\ref{related} lists the related works. 
And finally, Section~\ref{concl} concludes this paper.

%% file: background.tex
\section{Background}
\label{backgrnd}
This section introduces terms and concepts related to our SLR on mobile app security in developing countries.

\subsection{Mobile app security}
Mobile app security pertains to the measures implemented to safeguard mobile apps from various forms of threats, including hacking, mobile malware, data breaches, privacy violations, and other malicious activities\footnote{\url{https://fraudwatch.com/blog/what-is-mobile-app-security-including-8-application-security-tips/}}.
In practical terms, mobile app security includes various aspects integrated during the app design process, such as:
\begin{itemize}
    \item Ensuring that only authorized users can access the app by verifying their identity (e.g., using passwords, biometrics, or two-factor authentication).
    \item Controlling users' actions within the app based on their roles and permissions.
    \item Ensuring APIs used by the app are secure (e.g., validating input, using tokens). 
    \item Protecting against API abuse and unauthorized access.
    \item Avoiding hardcoding secrets in the source code.
    \item Securing data transmitted between the app and servers using secure protocols such as HTTPS.
    \item  Encrypting sensitive data when storing on the device.
    \item Storing sensitive data (such as passwords, tokens, or keys) securely within the app
    \item Implementing runtime security controls (e.g., obfuscation, anti-tampering mechanisms) to prevent reverse engineering and code modification.
    \item Keeping the app up-to-date with security patches and bug fixes.
\end{itemize}

Ignoring these aspects can lead to various security risks, including data breaches, privacy violations, and mobile malware attacks.
Specifically, a data breach in the context of mobile apps refers to a security incident where unauthorized parties gain access to sensitive or confidential information stored within the app, including personally identifiable information, credit card number, and medical records~\cite{proofpoint2024}. As for privacy violations in mobile apps, it refer to instances where an app mishandles sensitive user data, potentially compromising user privacy. Besides, mobile malware is malicious software that can exploit vulnerabilities to compromise the user's data security and privacy from the device or other installed apps. As for vulnerable mobile apps, they are apps that contain exploitable risks and unsecured entry points that threat actors, including malware, can leverage.

Given that mobile apps often handle sensitive information, such as financial records, personal health data, and location data.
Numerous security concerns, including insecure communication, insecure data storage, improper usage of cryptography, and improper platform usage~\cite{ValueM}, can compromise the security of this information, leading to identity theft, financial losses, or reputational damage. 

Adversaries can exploit these security issues using various methods, including Man-in-the-Middle attacks, authentication attacks, device theft, SMS interception, and more~\cite{Castle}.

%% file: methodology.tex
\section{Methodology}
\label{method}

\begin{figure}[!t]
    \centering
    \includegraphics[width=0.75\textwidth]{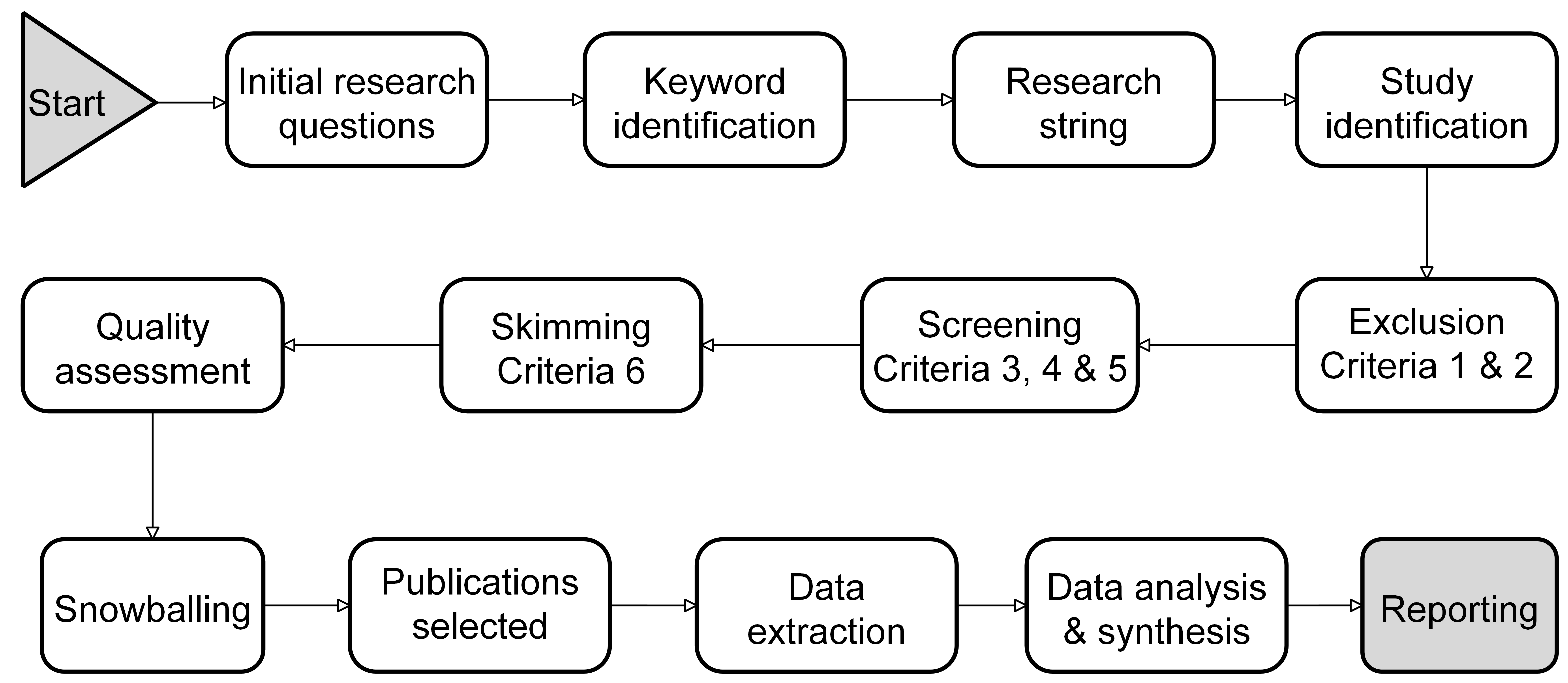}
    \caption{Overview of the methodology followed in this SLR}
    \label{methodology}
\end{figure}
In this study, we have followed the systematic literature review guideline proposed by Keele~\cite{keele2007guidelines}.

The research process, as depicted in Figure~\ref{methodology}, can be broken down as follows:
\begin{itemize}
    \item We initiated the process by formulating precise research questions, serving as a foundation for identifying relevant publications.
    \item Guided by our research questions, we identified pertinent keywords. 
    \item Previous keywords were instrumental in pinpointing a substantial body of relevant studies, culminating in constructing a comprehensive search string.
    \item Leveraging the search string developed in the prior step, we conducted systematic searches in four reputable digital libraries: IEEE Xplore, ACM Digital Library, Springer, and ScienceDirect.
    \item Having gathered publications from the digital libraries, we applied predefined exclusion criteria to filter out the most relevant publications.
    \item To further refine our selection, we meticulously reviewed the titles and abstracts of each paper, adhering to specific criteria. Only studies meeting these criteria were retained.
    \item We established a list of criteria to assess the quality of publications by removing pour-quality publications.
    \item We also employed backward snowballing of primary publications to identify any potentially missed relevant ones.
    \item We extracted data from selected publications, allowing us to answer research questions.
    \item Then, we based on the data extracted to perform data analysis to guide the interpretation of the results.
    \item The final step reports our findings to the research community.
\end{itemize}

\subsection{Initial research questions}
To provide an understanding of the publications produced by researchers to address the specific challenges of mobile app security, we have formulated a set of research questions (RQs). These RQs serve as the backbone of our investigation, and each sheds light on key aspects of this research field.

\textbf{RQ0: In which venues are these studies published?} Understanding where these studies are published provides a better understanding of the spread of knowledge about mobile app security in developing countries. This question overviews the conferences and journals where researchers have published their work. 

\textbf{RQ1: What are the research directions around mobile app security in developing countries?} Mobile app security research can take various directions. Investigating these directions in developing countries points out the areas where researchers have chosen to concentrate their efforts. This knowledge informs us about the prevailing concerns and priorities in this context.

\textbf{RQ2: What are the security concerns addressed in the literature?} The security of mobile apps is a fundamental concern worldwide. As the use of mobile apps continues to grow in developing countries, understanding the specific security issues that researchers address becomes crucial. This question unveils the most frequent security concerns relevant to mobile apps in these regions.

\textbf{RQ3: Which apps are covered in the literature?} This research question enables us to identify which of the multitude of application categories available are covered in the literature for developing countries. Recognizing the categories publications have focused on helps identify the gaps and areas needing further exploration.

\textbf{RQ4: What techniques are used to detect security issues?} Researchers employ various techniques to detect security issues in mobile apps. By unveiling these techniques, we gain insights into the strategies employed to identify security issues in the context of developing countries.

\textbf{RQ5: What motivated researchers to investigate mobile app security in developing countries?} This question elucidates the various motivations mentioned in the literature, providing an understanding of the factors that have inspired researchers to investigate mobile app security in developing countries.

Each research question plays a key role in clarifying the multifaceted landscape of mobile app security in developing countries, contributing to a more informed and comprehensive understanding of this critical field.

\subsection{Search strategy}
Getting relevant publications began by crafting a search string and executing our search across four well-regarded repositories.

\textit{Search string}.
Our search string was constructed based on a list of keywords derived from our previous research questions. These keywords were categorized into three groups, as displayed in Table~\ref{table:keyword}: "Device," "Security," and "Target." Each line in the table represents a category with multiple corresponding keywords. For each category, we created a string, denoted as $c_i$ (with $i$ representing the category number), using a disjunction of its keywords, such as $c_1$ = $k_1$ OR $k_2$ OR $k_3$ OR … OR $k_n$. We obtained three strings ($c_1$, $c_2$, $c_3$). We combined these three strings in our search string as a conjunction, resulting in the final search string, i.e., final\_string = $c_1$ AND $c_2$ AND $c_3$. This final string was used in online repositories for our systematic search.
\begin{table}[htb!]
    \centering
   \caption{Keywords used in this SLR.}
   \label{table:keyword}
\begin{adjustbox}{width=.9\linewidth,center}

\begin{tabular}{|@{}c|l@{}|} 
\hline
\textbf{Category}&\textbf{Search keywords} \\
            \hline
             & android; mobile; smartphone; "smart phone"; iphone; ios; \\
            Device& "portable device"; app* \\
            \hline
              & security; malware; vulnerabilit*; weak*; exploit; flaw; breach; leak*; malicious; \\
             Security & phishing; ransomware; trojan; attack; compliance; crypto*; forensic; \\ & "reverse engineering"; encryption; threat; hack*; "privacy violation" \\
            \hline
              & "developing countr*"; "developing world"; "low-income countr*"; \\ Target & "middle-income countr*"; "low and middle-income countr*"; Africa*; \\ & "south Asia*"; "least-developed countr*"; \textbf{Country names}\tablefootnote{\scriptsize {Egypt; Libya; Angola; Benin; Botswana; "Burkina Faso"; Burundi; Cameroon; "Cabo Verde"; "Central African Republic"; Chad; Comoros; Congo; "Democratic Republic of the Congo"; "Ivory coast"; Djibouti; "Equatorial Guinea"; Eritrea; Eswatini; Ethiopia; Gabon; Gambia; Ghana; Guinea; "Guinea Bissau"; Kenya; Lesotho; Liberia; Madagascar; Malawi; Mali; Mauritania; Mauritius; Mozambique; Namibia; Niger; Nigeria; Rwanda; "St. Helena"; "Sao Tome \& Principe"; Senegal; "Sierra Leone"; Somalia; "South Sudan"; Sudan; Tanzania; Togo; Uganda; Zambia; Zimbabwe; Albania; Armenia; Azerbaijan; Belarus; "Bosnia \& Herzegovina"; Georgia; Kosovo; Macedonia; Moldova; Montenegro; Serbia; Ukraine; Belize; "Costa Rica"; Cuba; Dominica; "Dominican Republic"; "El Salvador"; Grenada; Guatemala; Haiti; Honduras; Jamaica; Montserrat; Nicaragua; Panama; "St. Lucia"; "St. Vincent and the Grenadines"; Bolivia; Ecuador; Guyana; Paraguay; Peru; Suriname; Venezuela; Afghanistan; Bhutan; Cambodia; Kyrgyzstan; "Lao People’s Democratic Republic"; Maldives; Mongolia; Myanmar; Nepal; Pakistan; Tajikistan; "Timor Leste"; Turkmenistan; Uzbekistan; Jordan; Lebanon; "Syrian Arab Republic"; "West Bank and Gaza Strip"; Yemen; "Cook Islands"; Fiji; Kiribati; "Marshall Islands"; Micronesia; Nauru; Niue; Palau; "Papua New Guinea"; Samoa; "Solomon Islands" ; Tokelau; Tonga; Tuvalu; Vanuatu; "Wallis \& Futuna"; Philippines; Morocco; Bangladesh; India}} \\
 \hline
\end{tabular}
\end{adjustbox}
\end{table}

\textit{Well-known repositories}.
We conducted our SLR across four repositories: IEEE Xplore, ACM Digital Library, ScienceDirect, and Springer.
The choice of these repositories is motivated by the fact that they are the most important and widely used in the scientific community.
We employed Advanced Search for the first three repositories to refine our results, focusing on each publication title and abstract. 
In Springer, we cannot focus our search on the title or the abstract; we performed a classic search across all metadata and filtered the results to computer science publications.
To account for specific constraints related to search fields, we subdivided the main search string into multiple strings before executing the search. Notably, IEEE Xplore allows a maximum of 25 keywords in one clause. ScienceDirect restricts string searches to a maximum of 8 boolean operators. While ACM Digital Library and Springer did not explicitly describe these constraints in their documentation, we opted to apply the same limitations as IEEE Xplore to ensure optimal results.

\textit{Handling large results}.
Situations occur where our searches returned more than 1,000 items, exceeding Springer's display limit. 
We employed a Python script to scrape all findings. 
Additionally, Python script-based verification was used to verify search strings within the title and abstract of publications retrieved from Springer and ScienceDirect, ensuring consistency in the filtering process. 
Unlike IEEE Xplore and ACM DL, ScienceDirect's search may include publications that do not contain the complete search string in the title or abstract, 
This also requires the Python script-based verification for additional checking to filter results effectively.

By employing these methods, we ensured comprehensive coverage and accuracy in our search across these repositories.
\subsection{Exclusion criteria}
\begin{figure}[!t]
    \centering
    \includegraphics[width=0.75\textwidth]{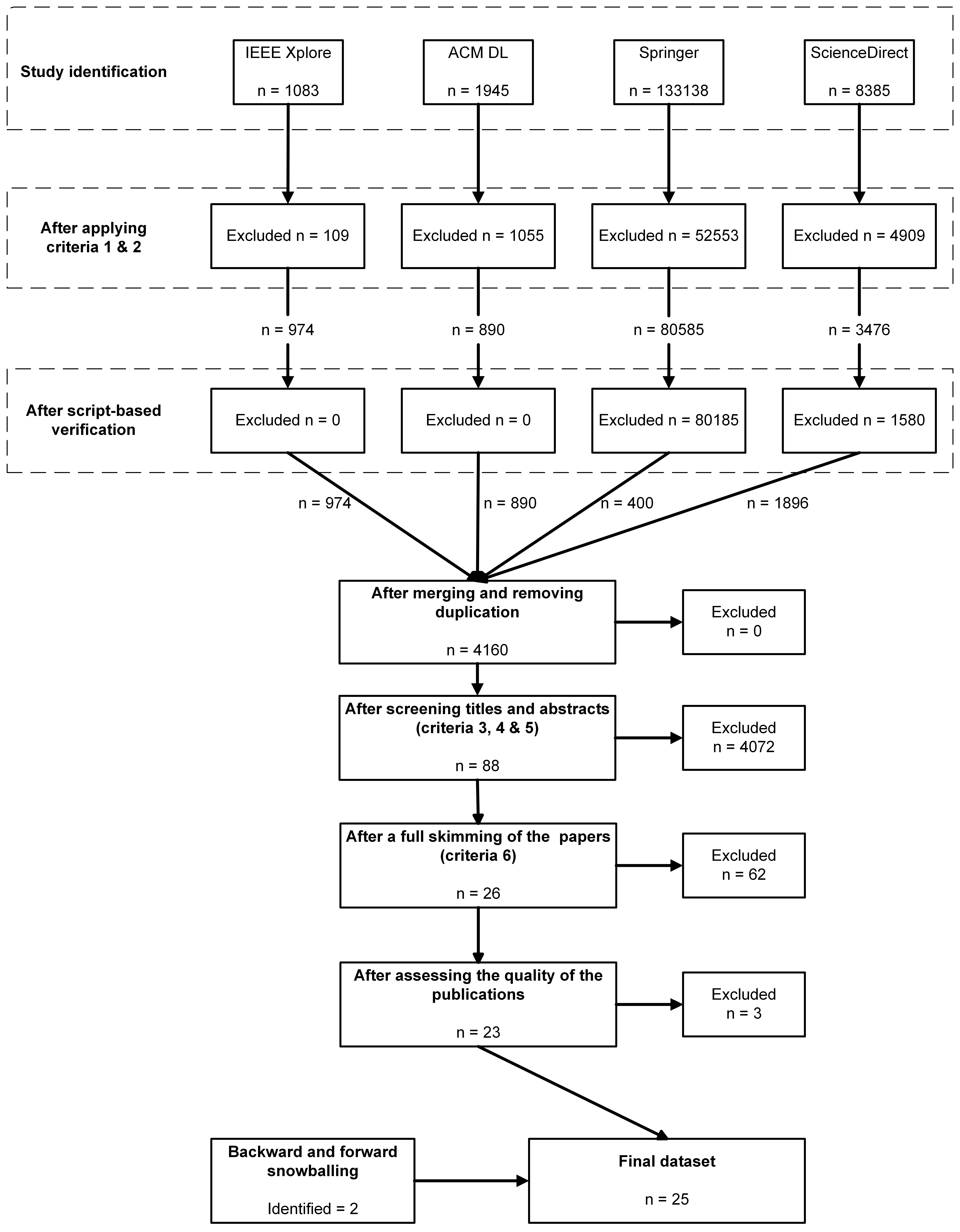}
    \caption{Process of publication selection.}
    \label{selection}
\end{figure}
Following our search across cited repositories, we encountered a multitude of publications, including many irrelevant to our SLR due to the use of generic keywords. To curate a consistent dataset of relevant publications, we applied a set of exclusion criteria:
\begin{enumerate}
    \item[1.] \textbf{Remove duplicate publications.} We employed a Python script to identify and remove duplicated entries based on title, abstract, author names, and year of publication. 
    We found around \num{58212} of duplicated publications that we removed.
    \item[2.] \textbf{Exclude books and thesis reports.} Our search on ACM Digital Library returned several books and thesis reports that were not aligned with our SLR objectives. Therefore, we removed these entries around 414), focusing on journal articles and conference papers.
    \item[3.] \textbf{Filter out unrelated publications}. Some keywords in our search string yielded papers unrelated to mobile apps. We manually reviewed the remaining publications' titles and abstracts, excluding those deemed irrelevant. This step further refined our dataset, discarding approximately 98\% of the results.
    \item[4.] \textbf{Exclude non-security-focused publications.} Our dataset contains publications exclusively discussing mobile apps without addressing security. During the manual review of publications' titles and abstracts, we removed those publications.
    \item[5.] \textbf{Remove non-developing-country papers.} We exclude publications not targeting developing countries during the manual review. After this step, we have 88 remaining publications.
    \item[6.] \textbf{Comprehensively review each paper.} We ensured each publication explicitly focuses on mobile application security in developing countries by reviewing each paper entirely. We excluded 62 irrelevant publications.
\end{enumerate}

By implementing these exclusion criteria, we systematically filtered and refined our dataset, ensuring that the publications included in our SLR were directly related to mobile app security in developing countries. 
At this stage, our dataset contains 26 publications.
Figure~\ref{selection} presents an overview and offers a clearer picture of the publication selection process.

\subsection{Quality assessment}
Given that there is no universally accepted definition for ‘quality’ in the context of a study, the assessment of a study’s quality can vary significantly depending on the purpose of the study~\cite{kitchenham2009systematic, yang2021quality}.
The quality assessment process serves as an additional layer of scrutiny after applying general exclusion criteria. This step is crucial in mitigating biases that may arise from studies of low quality. 
In our study, we perform the quality assessment to provide a better quality of publications.
This enables us to address the research questions more effectively and guides the analysis and interpretation of the results.
Based on~\cite{kitchenham2009systematic, yang2021quality}, we have defined 5 quality assessment (QA) criteria that help us ensure optimal publication quality.
These criteria are detailed in Table~\ref{table:quality_assessment}.
We restrict our consideration to those publications that fulfill at least 3 of these QA criteria.
As a result of this selection process, we have excluded 3 from the 26 remaining publications.
These 3 papers indeed only fulfill 2 of the criteria listed in Table~\ref{table:quality_assessment}.
\begin{table}[hbt!]
    \centering
    \caption{Criteria used for assessing the quality of the publications.}
    \label{table:quality_assessment}
    \begin{adjustbox}{width=\linewidth,center}
          \begin{tabular}{c|l}
            &Assessment criteria\\
            \hline
              QA1&Is the paper clearly motivated? \\
              QA2&Are the objectives of the paper clearly stated?  \\
              QA3&Does the paper provide a detailed description of the procedures used? \\
              QA4&Are the results of the paper clearly presented and interpreted in the context of the objectives? \\
              QA5&Does the paper discuss the key contributions?  \\
            \hline
        \end{tabular}
    \end{adjustbox}
\end{table} 

\subsection{Backward snowballing}
Since we have focused on the most important and widely used repositories in the scientific community, we could probably miss relevant publications published in other repositories. To deal with that, we initiated a backward and a forward snowballing process once we identified the primary publications relevant to our SLR. This involved a combination of automated scripts and manual efforts to collect references from these publications (backward snowballing) and publications having cited our dataset publications (forward snowballing). Using these collected publications, we utilized scripts to verify and remove any publications that did not align with the focus of our SLR.

Following this initial screening, we manually reviewed the remaining publications. If we encountered relevant publications that were not part of our originally selected primary studies, we incorporated them into our dataset. This iterative process ensured the comprehensive inclusion of pertinent literature in our SLR.
We retrieved two additional publications from the backward snowballing.

Table \ref{tab:full_list} presents the full list of the selected paper.
\begin{table*}[t]
  \caption{The full list of the primary publications selected.}
  \label{tab:full_list}
  \begin{adjustbox}{width=\linewidth,center}
  \begin{tabular}{clll}
    \toprule
    \textbf{Year}&\textbf{Venue type}&\textbf{Venue}&\textbf{Publication title}\\
    \midrule
    2023&Conference&ICOEI&Effective Security Testing of Mobile Applications for Building Trust in the Digital World~\cite{10125814}\\
    2023&Conference&ACM SIGCAS/SIGCHI&Evaluating Mobile Banking Application Security Posture Using the OWASP’s MASVS Framework~\cite{chiboora2023evaluating}\\
    2022&Conference&IEEE S\&P&“Desperate Times Call for Desperate Measures”: User Concerns with Mobile Loan Apps in\\
    &&&Kenya \cite{Munyendo}\\
    2022&Journal&HPT&Adoption of Covid-19 contact tracing app by extending UTAUT theory: Perceived disease threat \\
    &&&as moderator \cite{CHOPDAR2022100651}\\
    2022&Conference&ICKES&User's Perception on Security and Privacy in Using Crypto Currency Trading Application in\\
    &&&India \cite{10060666}\\
    2022&Conference&ICCCNT&State of Survey: Advancement of Knowledge Environmental Sustainability in Practicing\\
    &&&Administrative Apps \cite{9984416}\\
    2021&Journal&HPT&Understanding digital contact tracing app continuance: Insights from India \cite{PRAKASH2021100573}\\
    2021&Journal&RCS&Determining factors and impacts of the intention to adopt mobile banking app in Cameroon: \\
    &&& Case of SARA by afriland First Bank \cite{Kamdjoug}\\
    2021&Conference&ICSCCC&Security Issues of Unified Payments Interface and Challenges: Case Study \cite{9478078}\\
    2021&Conference&CT-RSA&Mesh Messaging in Large-Scale Protests: Breaking Bridgefy \cite{10.1007/978-3-030-75539-3_16}\\
    2020&Conference&AFRICOMM&Analysis of the Impact of Permissions on the Vulnerability of Mobile Applications \cite{Koala}\\
    2020&Conference&InterSol&Vulnerability Analysis in Mobile Banking and Payment Applications on Android in African\\
    &&&Countries \cite{Bassole}\\
    2020&Conference&COMPASS&We Don't Give a Second Thought Before Providing Our Information: Understanding Users' \\
    &&&Perceptions of Information Collection by Apps in Urban Bangladesh \cite{10.1145/3378393.3402244}\\
    2020&Conference&COMS2&Signature Based Malicious Behavior Detection in Android \cite{10.1007/978-981-15-6648-6_20}\\
    2020&Conference&USENIX Security&Security Analysis of Unified Payments Interface and Payment Apps in India \cite{247668}\\
    2019&Conference&ICSIoT&A Comparative Study of User Data Security and Privacy in Native and Cross Platform Android\\
    &&&Mobile Banking Applications \cite{Ansong}\\
    2019&Conference&ICCSA&Forensic Analysis of Mobile Banking Apps \cite{Osho}\\
    2019&Journal&MAT&Forensic analysis of mobile banking applications in Nigeria \cite{Uduimoh}\\
    2018&Conference&RTEICT&Integrating OAuth and Aadhaar with e-Health care System \cite{9012487}\\
    2017&Conference&ICTD&A Study of Static Analysis Tools to Detect Vulnerabilities of Branchless Banking Applications in\\
    &&&Developing Countries \cite{Ibrar}\\
    2017&Journal&TOPS&Mo(Bile) Money, Mo(Bile) Problems: Analysis of Branchless Banking Applications \cite{Reaves}\\
    2017&Conference&ICEEG&Side-Effects of Permissions Requested by Mobile Banking on Android Platform: A Case Study of\\
    &&&Morocco \cite{10.1145/3108421.3108433}\\
    2017&Conference&NSysS&Vulnerability detection in recent Android apps: An empirical study \cite{7885802}\\
    2016&Conference&ANTS&On the MitM vulnerability in mobile banking applications for android devices \cite{7947811} \\
    2016&Conference&ACM DEV&Let's Talk Money: Evaluating the Security Challenges of Mobile Money in the Developing World \cite{Castle}\\
  \bottomrule
\end{tabular}
\end{adjustbox}
\end{table*}

\subsection{Data extraction}
\begin{table*}[t]
  \caption{The list of the data extracted in the selected publications.}
  \label{tab:data_extraction}
  \begin{adjustbox}{width=\linewidth,center}
  \begin{tabular}{lll}
    \toprule
    \textbf{Extracted information}&\textbf{Description}&\textbf{Corresponding RQ}\\
    \midrule
    Venue name&The name of the venue where the paper has been published.&\\
    Venue Type&The type of the venue where the paper has been published.&RQ0\\
    Venue location&The location where the venue held.&\\
    Author institution&The origin of the publication (the affiliations of the authors).&\\
    \midrule
    Type of study&The type of contribution researchers performed.&RQ1\\
    \midrule
    Secuirty issues&The issues of security researchers addressed in their publications.&RQ2\\
    \midrule
    Category of app&The categories (according to GooglePlay) of app researchers studied in their works.&RQ3\\
    \midrule
    Issue detection technique&The techniques and methods used to detect issues on the apps.&RQ4\\
  \bottomrule
\end{tabular}
\end{adjustbox}
\end{table*}

In this section, we present the information extracted to address the research questions mentioned earlier.
Table~\ref{tab:data_extraction} provides an overview of the data extracted from the selected publications. 
To answer RQ0, we collected metadata for each publication from the repository websites where they were published.
These repositories offer comprehensive information about the publications. The extracted data include details related to the venues and the authors’ affiliations.
Next, we thoroughly read each publication to extract relevant information. Specifically:
\begin{itemize}
    \item For RQ1, we identified the types of studies conducted by researchers.
    \item To address RQ2, we examined the security concerns they tackled.
    \item For RQ3, we categorized the apps they focused on.
    \item Finally, to answer RQ4, we analyzed the various techniques used to detect and address security issues in these apps. 
\end{itemize}
We have decided to group the selected publications into types of study after reading and getting results from the extraction of data.
More details have been provided about these extracted data in the next section.

\input{data_analysis}

%% file: data_analysis.tex
\subsection{Data analysis and synthesis}
\label{data_analysis}
\begin{figure}[h!]
  \centering
  \includegraphics[width=\linewidth]{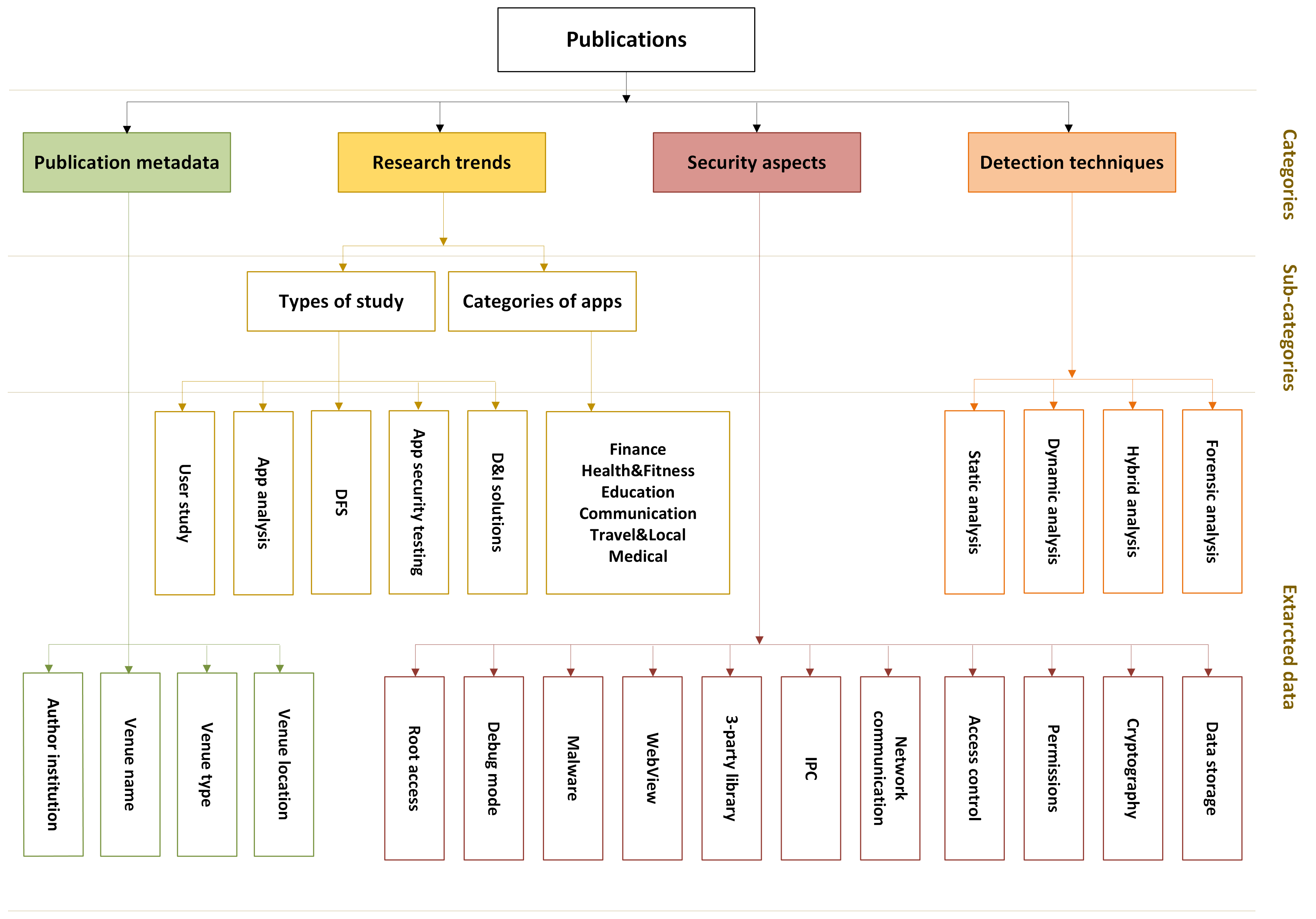}
  \caption{Overview of information extracted from the publications}
  \label{taxonomy}
\end{figure}
In the previous section, we extracted several data to address research questions.
 After thoroughly reviewing the publications and collecting relevant information, we categorized the information into four main groups, as depicted in Fig.~\ref{taxonomy}.

\textbf{Publication metadata (RQ0).} The extracted metadata provides details about the publication venue (conference or journal), venue name, location (for conferences), and author affiliations. 
We present these results in a comprehensive table (see Table~\ref{tab:rq0_2}).

\begin{figure}[t!]
    \centering
   \begin{minipage}{0.48\textwidth}
     \centering
   \includegraphics[width=1.1\linewidth]{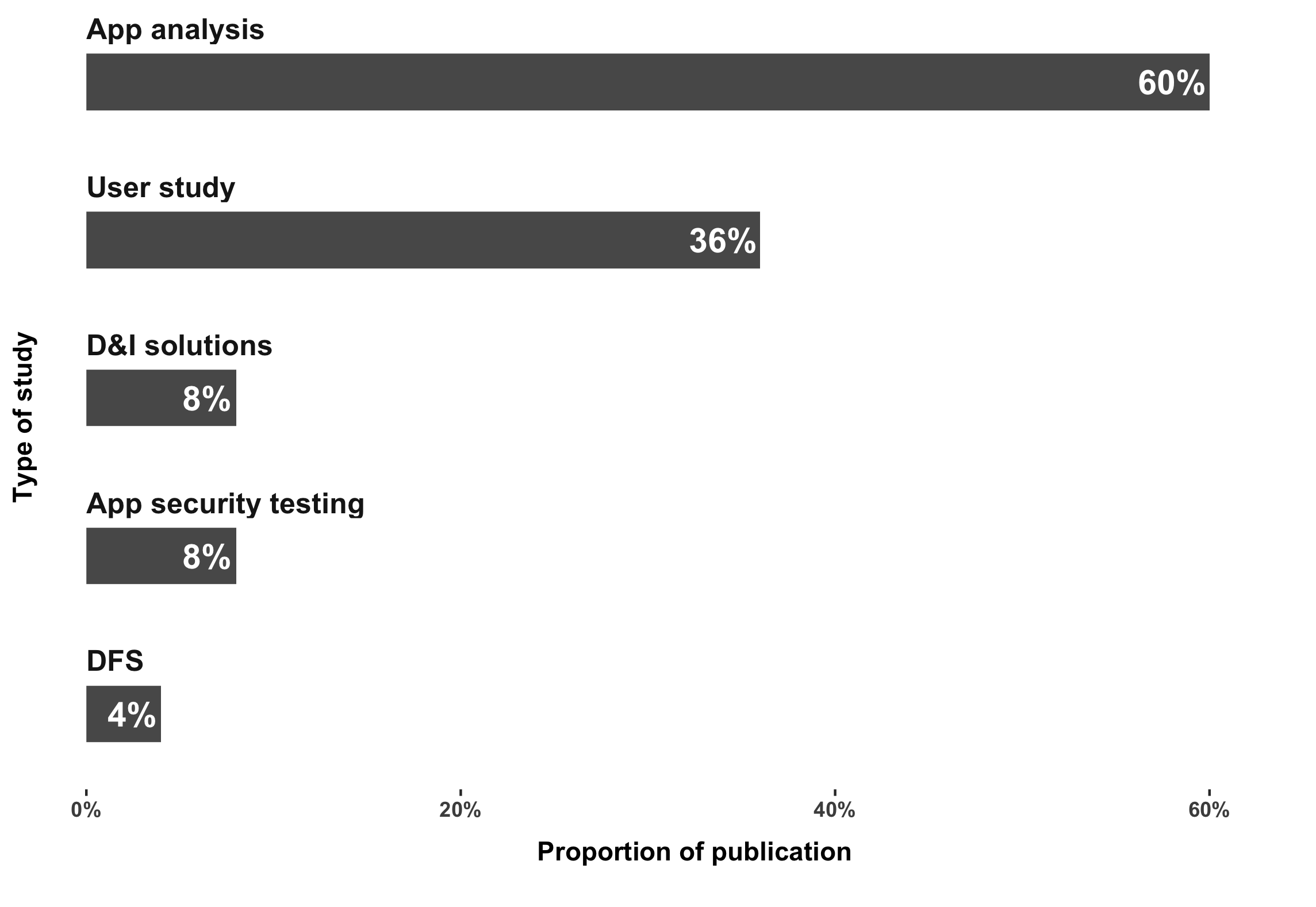}
    \caption{Proportion of publications per type of study.}
    \label{fig:type}
   \end{minipage}\hfill
   \begin{minipage}{0.48\textwidth}
     \centering
     \includegraphics[width=.95\linewidth]{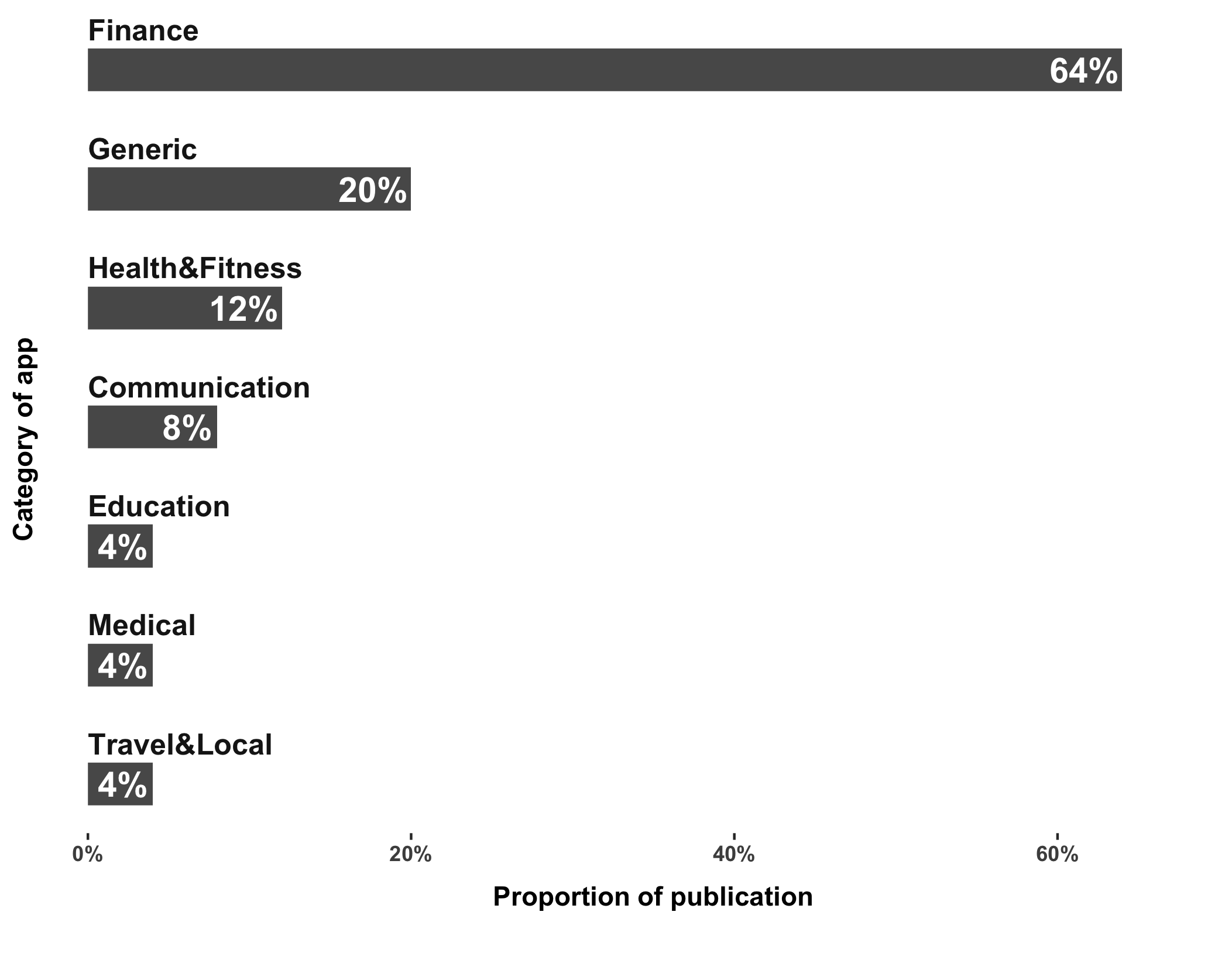}
    \caption{Proportion of publications per category of app.}
    \label{fig:cat}
   \end{minipage}
\end{figure}

\textbf{Research trends (RQ1 and RQ3).} For RQ1, we grouped the selected publications into five distinct types of study: user study, app analysis, development framework study (DFS), app security testing,  and study designing and implementing solutions (D\&I solutions).
Fig.~\ref{fig:type} illustrates the distribution across these study types, with app analysis and user study being the most common (60\% and 36\%, respectively).
We provide a summary, the strengths, and the limitations of each study type in Appendix~\ref{app}. 
For RQ3, the data extracted revealed various app categories addressed by researchers, including finance, health\&fitness, education, communication, travel\&local, and medical apps.
Fig.~\ref{fig:cat} displays the number of publications focused on each category. 
Financial apps received significant attention (around 64\% of the publications), followed by health\&fitness (12\%) and communication apps (8\%).
Some studies did not specify the app category, which we grouped under “Generic”, representing 20\% of the publications.
Other categories addressed by researchers, such as education, travel\&local, and medical, each represented no more 4\% of the publications.

\begin{figure}[t!]
    \centering
   \begin{minipage}{0.48\textwidth}
     \centering
   \includegraphics[width=1.0\linewidth]{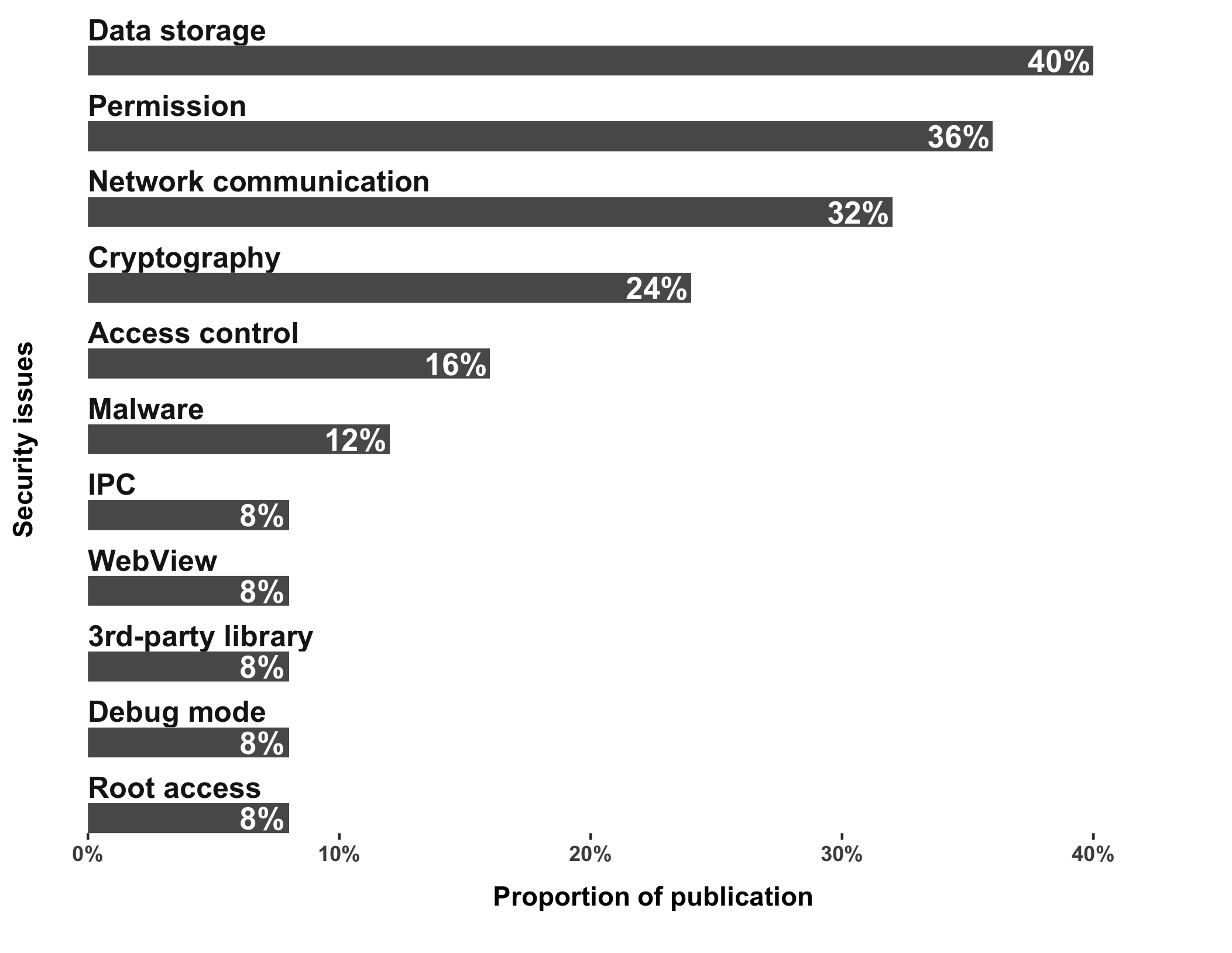}
    \caption{Proportion of publications per security issue.}
    \label{fig:issues}
   \end{minipage}\hfill
   \begin{minipage}{0.48\textwidth}
     \centering
     \includegraphics[width=1.0\linewidth]{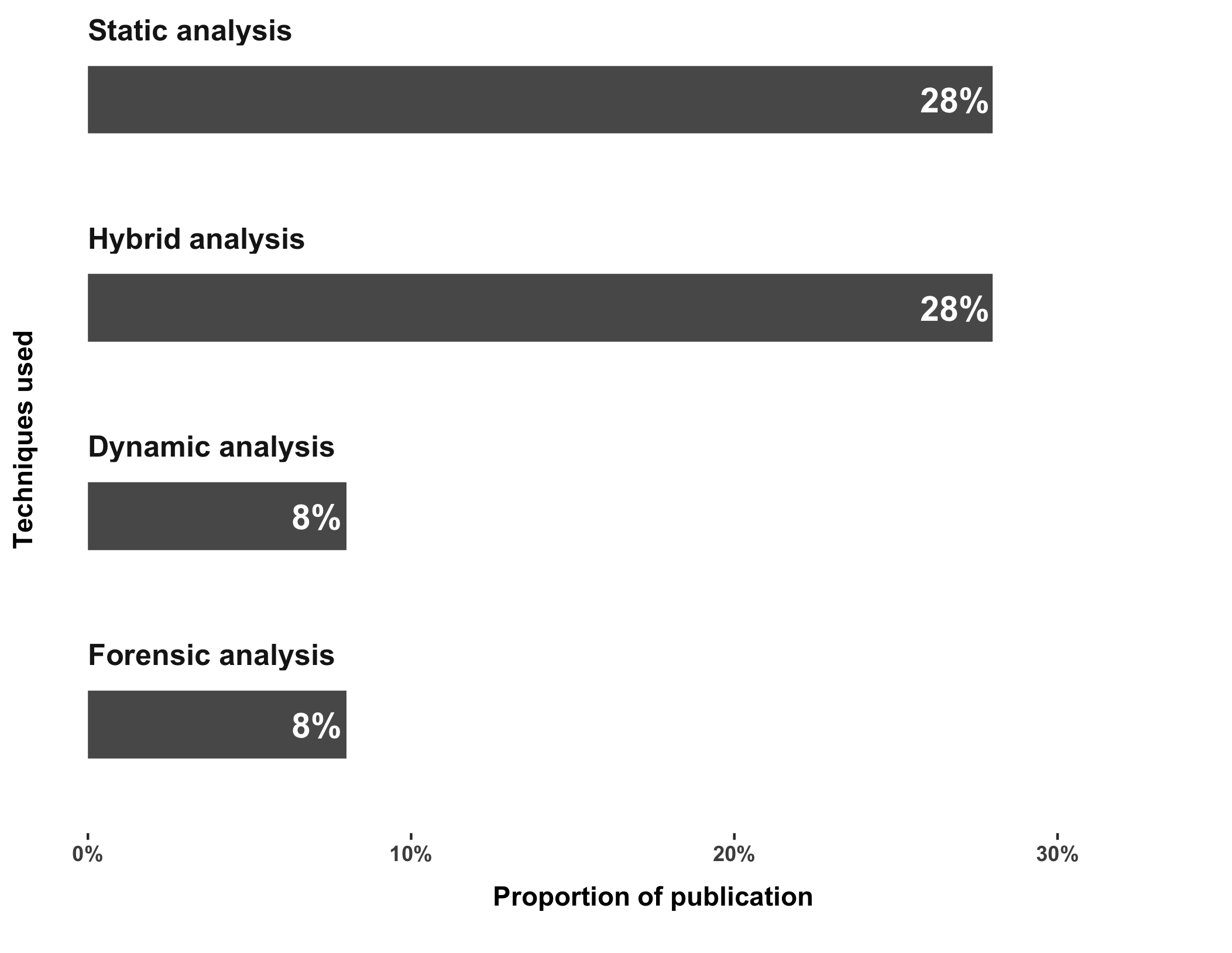}
    \caption{Proportion of publications per technique of detection.}
    \label{fig:technique}
   \end{minipage}
\end{figure}

\textbf{Security aspects (RQ2).} After reading each publication, we identified 11 security concerns addressed in the literature.
These include data storage, cryptography, permissions, access control, network communication, Inter-Process Communication (IPC), third-party libraries, webView, mobile malware, root access, and debug mode implementation.
Fig.~\ref{fig:issues} shows the number of publications addressing each concern, with data storage, permissions, network communication, and cryptography being the most common (40\%, 36\%, 32\%, and 24\%).

\textbf{Detection techniques (RQ4).} Researchers primarily employed four techniques.
Those techniques include static, dynamic, hybrid, and forensic analysis. 
Static analysis and hybrid (combining static and dynamic) analysis appear to be the most common techniques researchers use to address security issues, as illustrated in Fig.~\ref{fig:technique}.
\\

For a more detailed analysis and interpretation of the results, refer to Section~\ref{result}.

%% file: result.tex
\section{Results}
\label{result}
\begin{table}[t!]
  \caption{Mobile app origin area (\textbf{N.S}: Not Specified).}
  \label{tab:app_orig}
  \begin{adjustbox}{width=0.5\linewidth,center}
  \begin{tabular}{ccc}
    \toprule
    \textbf{Paper}&\textbf{App country origin}&\textbf{Contient}\\
    \midrule
    \cite{Castle} & N.S & Dev. countries \\
    \rowcolor{lightgray!50}\cite{Koala} & Burkina Faso & Africa \\
    \cite{Ibrar} & N.S & Dev. countries \\
    \rowcolor{lightgray!50}\cite{Bassole} & N.S & Africa \\
    \cite{Osho} & Nigeria & Africa \\
    \rowcolor{lightgray!50}\cite{10125814} & India & Asia \\
    \cite{chiboora2023evaluating} & N.S & Africa \\
    \rowcolor{lightgray!50}\cite{Munyendo} & Kenya & Africa \\
    \cite{CHOPDAR2022100651} & India & Asia \\
    \rowcolor{lightgray!50}\cite{10060666} & India & Asia \\
    \cite{9984416} & Bangladesh & Asia \\
    \rowcolor{lightgray!50}\cite{PRAKASH2021100573} & India & Asia \\
    \cite{Kamdjoug}  & Cameroon & Africa \\
    \rowcolor{lightgray!50}\cite{9478078} & India & Asia \\
    \cite{10.1007/978-3-030-75539-3_16} & Belarus, Zimbabwe & Africa, Europe \\
    \rowcolor{lightgray!50}\cite{10.1145/3378393.3402244} & Bangladesh & Asia \\
    \cite{10.1007/978-981-15-6648-6_20} & India & Asia \\
    \rowcolor{lightgray!50}\cite{247668} & India & Asia \\
    \cite{Ansong} & Ghana & Africa \\
    \rowcolor{lightgray!50}\cite{Uduimoh} & Nigeria & Africa \\
    \cite{9012487} & India & Asia \\
    \rowcolor{lightgray!50}\cite{Reaves} & N.S & Dev. countries \\
    \cite{10.1145/3108421.3108433} & Morocco & Africa \\
    \rowcolor{lightgray!50}\cite{7885802} & Bangladesh & Asia \\
    \cite{7947811} & India & Asia \\
  \bottomrule
\end{tabular}
\end{adjustbox}
\end{table}

In this section, we present and interpret the study results and answer the research questions.
The study consists of identifying research works concerning the security of mobile apps used in the context of developing countries.
We have identified 25 publications investigating mobile app security.
All have, indeed, investigated mobile apps used in developing countries.
In Table~\ref{tab:app_orig}, we map publications with the context of the study to show where the apps addressed are used.

\subsection{Publication venue and paper origin} 
This section details the distribution of publications according to venue type, geographic origin, and conference locations.
In this study, a paper is considered to originate from a specific place if all authors' affiliations match that location. 

We have found that 81.5\% of the publications are presented at conferences, with the remaining 18.5\% in journals. Almost every venue accounts for one publication, except one journal venue for two.

Given the preponderance of conference publications in the literature, it is noteworthy that researchers often present their work at conferences held in the country of their institutions or in a developing country. 
Indeed, approximately 77\% of conference publications (corresponding to 17) originate from developing country institutions. Among these publications, four~\cite{Osho, 10.1145/3108421.3108433, 8701396, 9988647} were not presented at conferences held in developing countries, as illustrated in Table~\ref{tab:rq0_2}.
Conversely, non-developing countries contribute a few conference publications~\cite{10.1007/978-3-030-75539-3_16, 10.1145/3378393.3402244, Castle, 247668, Munyendo}, with one taking place in a developing country~\cite{Castle}.

Among the journal publications, one is identified as non-developing countries' contribution~\cite{Reaves}.
\\

\begin{adjustbox}{width=\linewidth}
    \fcolorbox{black}{lightgray!50}{
        \begin{varwidth} {\linewidth}
            \textbf{RQ0:} Most studies have been presented at conferences held in developing countries. Nevertheless, 
            mobile app security in developing countries has attracted the interest of developers located outside these countries.
        \end{varwidth}
    }
\end{adjustbox}
\begin{table}[t!h]
  \caption{Overview of publications and venues.}
  \label{tab:rq0_2}
  \begin{adjustbox}{width=\linewidth,center}
  \begin{threeparttable}
   \begin{tabular}{lccclcl}
            \toprule
            {\Large \textbf{Date [paper]}}&{\Large \textbf{Venue}}&{\Large \textbf{Location}}&{\Large \textbf{VHIDC\tnote{*}}}&{\Large \textbf{1st author's institution}}&{\Large \textbf{ILIDC\tnote{**}}}&{\Large \textbf{Other authors' institutions (\# of authors)}}\\
            \hline\hline
             & & & \\
            {\LARGE Conference} & \\
            \hline
             2019 \cite{Ansong} & ICSIoT & Accra, Ghana & Yes & Kwame Nkrumah Univ. of Sci. and Tech., Kumasi, Ghana & Yes & Kwame Nkrumah Univ. of Sci. and Tech., Kumasi, Ghana (1) \\
             \midrule
             2017 \cite{Ibrar} & ICTD & Lahore, Pakistan & Yes & Information Technology University, Pakistan & Yes & Information Technology University, Pakistan (2) \\
               & & & & & & University of Washington, USA (1) \\
               \midrule
             2019 \cite{Osho} & ICCSA & Saint Petersburg, & No & Federal University of Technology, Minna, Nigeria & Yes & Federal University of Technology, Minna, Nigeria (3) \\
             & & Russia & & & & Covenant University, Ota, Nigeria (1) \\
             \midrule
             2016 \cite{Castle} & ACM DEV & Nairobi, Kenya & Yes & University of Washington, USA & No & University of Washington, USA (3)\\
             & &  & & & & Cornell University, New York, USA (1)\\
              \midrule
              & & & & & & University of Dhaka, Bangladesh (1) \\
             2020 \cite{10.1145/3378393.3402244} & COMPASS & Ecuador & No & Utah State University, USA & No & Jadavpur University, India (1) \\
              & & & & & & Utah State University, USA (1) \\
              & & & & &  &University of Toronto, Canada (2) \\
              \midrule
             2020 \cite{Bassole} & InterSol & Nairobi, Kenya & Yes & Université Joseph Ki-Zerbo, Ouagadougou, Burkina Faso & Yes & Université Joseph Ki-Zerbo, Ouagadougou, Burkina Faso (3) \\
              \midrule
             2022 \cite{Munyendo} & IEEE S\&P & San Francisco, CA, USA & No & Geoge Washington University, Washington, DC, USA & No & Geoge Washington University, Washington, DC, USA (2) \\
              \midrule
             2019 \cite{Koala} & AFRICOMM & Porto-Novo, Benin & Yes & Université Joseph Ki-Zerbo, Ouagadougou, Burkina Faso & Yes & Université Joseph Ki-Zerbo, Ouagadougou, Burkina Faso (4) \\
             \midrule
             2017 \cite{10.1145/3108421.3108433} & ICEEG & Turku, Finland & No & Dept. of Computer Sciences, University Cadi Ayyad, Marrakesh, Morocco & Yes & Dept. of Computer Sciences, University Cadi Ayyad, Marrakesh, Morocco (2)  \\   
             \midrule
             2022 \cite{10060666} & ICKES & Chickballapur, India & Yes & School of CSIT, Symbiosis University, Indore, India & Yes & School of CSIT, Symbiosis University, Indore, India (2) \\
             \midrule
             2016 \cite{7947811} & ANTS & Bangalore, India & Yes & SCIS, University of Hyderabad, Hyderabad, India & Yes & Centre for Mobile Banking, IDRBT, Hyderabad, India (2) \\
             \midrule
             2017 \cite{7885802} & NSysS & Dhaka, Bangladesh & Yes & Department of Computer Science and Engineering, Bangladesh University & Yes & Department of Computer Science and Engineering, Bangladesh University \\
             & & & & of Engineering and Technology & &of Engineering and Technology (2)  \\
             \midrule
             2022 \cite{9984416} & ICCCNT & Kharagpur, India & Yes & Dept. of CSE, Daffodil International University, Dhaka, Bangladesh & Yes & Dept. of CSE, Daffodil International University, Dhaka, Bangladesh (4) \\
             \midrule
             2023 \cite{10125814} & ICOEI & Tirunelveli, India & Yes & Centre for Development of Advanced Computing (C-DAC), Mumbai, India & Yes & Centre for Development of Advanced Computing (C-DAC), Mumbai, India (2) \\
             & & & & & & Ministry of Electronics and Information Technology (MeitY), Delhi, India (1) \\
             \midrule
             2023~\cite{chiboora2023evaluating} & ACM SIGCAS/SIGCHI & Cape Town, South Africa & No & Cylab-Africa/Upanzi Network Kigali, Rwanda & Yes & Cylab-Africa/Upanzi Network Kigali, Rwanda (3) \\
             \midrule
             2021 \cite{9478078} & ICSCCC & Jalandhar, India & Yes & Department of CSE, NIT, Karnataka, Surathkal, Mangalore, India & Yes & Department of CSE, NIT, Karnataka, Surathkal, Mangalore, India (2) \\
             \midrule
             2018 \cite{9012487} & RTEICT & Bangalore, India & Yes & Dept. of Computer Science and Engineering, B.M.S. College of Engineering, & Yes & Dept. of Computer Science and Engineering, B.M.S. College of Engineering, \\
             & & & & Bengaluru, India & & Bengaluru, India (1) \\
             \midrule
             2020 \cite{10.1007/978-981-15-6648-6_20} & COMS2 & Gujarat, India & Yes & Sardar Patel University of Police, Security and Criminal Justice, Jodhpur, India & Yes & Sardar Patel University of Police, Security and Criminal Justice, Jodhpur, India (1) \\
            & & & & & & National Institute of Technology, Raipur, Raipur, India (2) \\
             \midrule
             2021 \cite{10.1007/978-3-030-75539-3_16} & CT-RSA & Virtual Event & - & Royal Holloway, University of London, London, UK & No & Royal Holloway, University of London, London, UK (3) \\
             \midrule
             2020 \cite{247668} & USENIX Security & USA & No & University of Michigan & No & University of Michigan (3) \\
             \midrule
             {\large Total} & & & {\large 13} & & {\large 15} &\\
             \hline\hline
             & & & \\
            {\LARGE Journal} & \\
            \midrule
              & &  & &  & & University of Grenoble, Grenoble Cédex 9, France (1) \\
              2021 \cite{Kamdjoug} & RCS & - & - & Catholic University of Central Africa, Yaoundé, Cameroon & Yes & Catholic University of Central Africa, Yaoundé, Cameroon (1) \\
              & & & &  &  & Toulouse Business School, Toulouse, France (1) \\
              \midrule
              2017 \cite{Reaves} & TOPS & - & - & University of Florida, FL, USA & No & University of Florida, FL, USA (5) \\
              & & & & & & University of Illinois at urbana-champaign, USA (1) \\
              \midrule
              2019 \cite{Uduimoh} & MAT & - & - & Federal University of Technology, Minna, Nigeria & Yes & Federal University of Technology, Minna, Nigeria (2) \\
              & & & & & & Clemson University, Clemson, SC, USA (1) \\
              \midrule
              2022 \cite{CHOPDAR2022100651} & HPT & - & - & Indian Institute of Management, Shillong, India & Yes & - \\
              \midrule
              &  & & & Vinod Gupta School of Management, Indian Institute of Technology Kharagpur, &  & Vinod Gupta School of Management, Indian Institute of Technology Kharagpur,\\
              2021 \cite{PRAKASH2021100573} & HPT & - & - & West Bengal, India & Yes & West Bengal, India (1)\\
              & & & & & & Manipal Institute of Management, Manipal Academy of Higher Education, KA, India (1) \\
          \bottomrule
        \end{tabular}
        \begin{tablenotes}
            \item[*] \large Venue held in a developing country. 
            \item[**] Institution located in a developing country.
        \end{tablenotes}
\end{threeparttable}
\end{adjustbox}
\end{table}

\subsection{Types of studies}
\begin{table}[t!]
    \caption{Overview of the types of research.}
    \label{tab:rq1}
        \centering
        \begin{adjustbox}{width=.9\linewidth,center}
          \begin{tabular}{l||cccccc}
            \toprule
            Publication&User study&App analysis&DFS&App security testing&D\&I solution \\
            \midrule
             Albrecht et al. \cite{10.1007/978-3-030-75539-3_16} & & \XSolidBrush & & & \\
             \rowcolor{lightgray!50} Al-Ameen et al. \cite{10.1145/3378393.3402244} & \XSolidBrush & & & & \\
             Ansong et al. \cite{Ansong} & & &\XSolidBrush & & \\
             \rowcolor{lightgray!50} Bandan et al. \cite{9984416} & \XSolidBrush & & & & \\
             Bassolé et al. \cite{Bassole} & & \XSolidBrush & & & \\
             \rowcolor{lightgray!50} Castle et al. \cite{Castle} & \XSolidBrush & \XSolidBrush & & & \\
             Chiboora et al. \cite{chiboora2023evaluating} & \XSolidBrush & \XSolidBrush & & & \\
             \rowcolor{lightgray!50} Chopdar et al. \cite{CHOPDAR2022100651} & \XSolidBrush & & & & \\
             Ibrar et al. \cite{Ibrar} & & \XSolidBrush & & & \\
             \rowcolor{lightgray!50} Kaka et al. \cite{7947811} & & \XSolidBrush & & & \\
             Kamdjoug et al. \cite{Kamdjoug} & \XSolidBrush & & & & \\
             \rowcolor{lightgray!50} Kant Kamal et al. \cite{10125814} & & & & \XSolidBrush & \\
             Khatoon et al. \cite{9012487} & & & & & \XSolidBrush \\
             \rowcolor{lightgray!50} Koala et al. \cite{Koala} & & \XSolidBrush & & & \\
             Kumar et al. \cite{247668} & & \XSolidBrush & & & \\
             \rowcolor{lightgray!50} Latifa et al. \cite{10.1145/3108421.3108433} & & \XSolidBrush & & & \\
             Madwanna et al. \cite{9478078} & & \XSolidBrush & & & \\
             \rowcolor{lightgray!50} Munyendo et al. \cite{Munyendo} & \XSolidBrush & & & & \\
             Osho et al. \cite{Osho} & & \XSolidBrush & & & \\
             \rowcolor{lightgray!50} Prakash et al. \cite{PRAKASH2021100573} & \XSolidBrush & & & & \\
             Reave et al. \cite{Reaves} & & \XSolidBrush & & & \\
             \rowcolor{lightgray!50} Shezan et al. \cite{7885802} & & \XSolidBrush & & & \\
             Sihag et al. \cite{10.1007/978-981-15-6648-6_20} & & \XSolidBrush & & & \XSolidBrush \\
             \rowcolor{lightgray!50} Uduimoh et al. \cite{Uduimoh} & & \XSolidBrush & & & \\
             Vakare et al. \cite{10060666} & \XSolidBrush & & & \XSolidBrush & \\
             \midrule
             \textbf{Number of publication} & \textbf{9} &  \textbf{15} &  \textbf{1} &  \textbf{2} &  \textbf{2}\\
          \bottomrule 
        \end{tabular}
        \end{adjustbox}
\end{table}
This section outlines the types of study directions performed by researchers examining mobile app security in developing countries. 
As presented in Table~\ref{tab:rq1}, the literature primarily focuses on five main types: user studies, development framework study (DFS), app analysis, app security testing, and designing \& implementing solutions (D\&I Solution).

\textit{User study.} 
Security's impact on the adoption of mobile applications in developing countries is a central concern. 
Around 32\% of publications utilize surveys and interviews to gauge this impact. Some of these studies do not exclusively focus on the security of mobile apps but explore broader factors influencing app adoption. 
For instance, researchers employed models such as the Unified Theory of Acceptance and Use of Technology (UTAUT), the Health Belief Model (HBM), and the Expectation-Confirmation Model (ECM) to understand the factors influencing the adoption and the continuance to adopt mobile health apps in India~\cite{CHOPDAR2022100651, PRAKASH2021100573}. 
Similarly, studies in Cameroon leveraged the UTAUT2 framework, the Technology Acceptance Model (TAM), and the Protection Motivation Theory (PMT) to examine the factors influencing users' decisions to adopt mobile banking apps~\cite{Kamdjoug}. The studies incorporate additional factors, including perceived security and privacy, ease of use, trust in technology, user satisfaction, and perceived usefulness. 
The  Partial Least Squares Structural Equation Modeling (PLS-SEM) is utilized to analyze user questionnaire responses.
These studies focused on specific app users and found the factors mentioned above significantly impact the adoption of mobile banking and health apps.
Other user study-focused publications concentrate specifically on mobile app security.
For example, authors investigated security and privacy perceptions related to cryptocurrency trading apps in India by using supervised machine learning algorithms (such as Random Forest and Logistic Regression) to analyze user reviews and sentiments~\cite{10060666}. In Kenya and Bangladesh, researchers explored user perceptions, behaviors, and privacy concerns associated with data collected by mobile apps~\cite{Munyendo, 10.1145/3378393.3402244}. The studies reveal that some users recognized the necessity of data collection for app functionality, while others expressed fear or indifference.

\textit{App analysis.} 
Most works performed in developing country contexts, approximately 60\%, are dedicated to detecting vulnerabilities and privacy violations through app analysis. 
App analysis involves examining various aspects of a mobile app to understand its performance, security, and user behavior.
A subset of the publications combines user studies with app analysis to measure the capacity of developers in implementing protection and ensuring users' security and privacy~\cite{Castle, chiboora2023evaluating}. 
However, the majority of the works have exclusively focused on app analysis, which predominantly centers on static, dynamic, forensic analysis, and a combination of static and dynamic analysis (see more details in Section~\ref{sub5}).
Despite the security challenges mobile applications face in developing countries~\cite{approov2023}, most publications do not use specialized approaches to deal with these challenges.
However, few studies have developed app analysis approaches specifically for the context of developing countries. 
For instance, knowing that developing countries face numerous threats in mobile financial services, Castle et al.~\cite{Castle} have focused on the features and development practices of mobile financial apps in this context. 
In other studies, authors have used approaches to specifically uncover vulnerabilities in UPI1.0 apps that banks from India used~\cite{247668, 9478078}. 

\textit{Development framework study.} 
This type of study delves into the usage of development platforms and their security guidelines in developing countries. 
Several studies show that Android apps often request more permissions than they need or use, impacting the security and privacy of the users' apps~\cite{khatoon2017privacy, alenezi2017abusing}.
Researchers, such as Ansong and Synaepa-Addision, think that stems from the development frameworks used~\cite{Ansong}. Aiming to ensure better user data privacy and security in mobile banking apps, they provide in their study a case in point, exploring mobile app security by comparing different development frameworks, including native and cross-platform, used to develop mobile banking apps in Ghana.
Development frameworks for developing mobile apps are not specific to the context of developing countries. Indeed, cross and native development frameworks are used worldwide.
Authors found that cross-platform apps tend to use more of the requested permissions compared to native apps.

\textit{App security testing.} 
App security analysis often involves penetration testing.
It simulates attacks on the mobile app to identify and address security vulnerabilities that could be exploitable, helping in enhancing the overall security posture of the app~\cite{bertolis}.
As they feel the need to address security issues on the apps, some researchers test the security posture of mobile apps from a specific AppStore in India's mSeva AppStore) using existing automated analysis tools for several security concerns~\cite{10125814}.
Their regressive testing methods focus on evaluating multiple facets of mobile apps before hosting in the Inda AppStore, emphasizing the critical need for robust security testing procedures. 
Similarly, others delve into India's mobile-based cryptocurrency trading apps' security and privacy aspects using penetration testing techniques~\cite{10060666}. 
In this work, authors use the OWASP mobile security testing framework to conduct a security analysis on the apps and identify several vulnerabilities.
However, all these security testing approaches are not specific to the unique context of developing countries.

\textit{Designing \& implementing solution.} 
Proposing sophisticated techniques is pivotal in fortifying the security of mobile apps worldwide. 
Some authors recognized this necessity by designing and implementing specialized solutions.
However, these solutions did not specifically target the concerns of developing countries.
Indeed, authors proposed a system that focuses on identifying malicious behaviors that could lead to several security concerns by analyzing system-generated logs during the execution of the app~\cite{10.1007/978-981-15-6648-6_20}.
The main goal is to provide a more dynamic and insightful detection method compared to existing techniques.
Besides, Khatoon and Umadevi have considered the problems facing developing country app users by creating a solution to ensure secure access, authentication, and authorization for Indian stakeholders within sensitive domains like healthcare~\cite{9012487}.
In their study, they integrate a known secure protocol for access delegation (OAuth 2.0) and the Aadhaar identification system of India into e-Health apps to enhance their security, offering secure access to health data and facilitating efficient interaction between patients and healthcare providers. The proposed system ensures the users' registration, login, and authentication and allows access without an Internet connection.
\\

\begin{adjustbox}{width=\linewidth}
    \fcolorbox{black}{lightgray!50}{
        \begin{varwidth} {\linewidth}
            \textbf{RQ1:} Our review highlights that the research on mobile app security in developing countries is broad, ranging from user studies to the design \& implementation of technical solutions. It is noteworthy, however, that most works have focused on app analysis.
        \end{varwidth}
    }
\end{adjustbox}
\subsection{Security concerns}
\begin{table}[t!]
\centering
  \caption{The security concerns addressed in the literature.}
  \label{tab:rq2}
  \begin{adjustbox}{width=\linewidth,center}
  \begin{tabular}{l||ccccccccccc}
    \toprule
    Publication&{Data Storage}&{Cryptography}&{Permission}&{Access Control}&{Network Comm.}&{IPC}&{3rd-party Library}&{WebView}&{Malware}&{Code Quality and Build}&{Resilience}\\
    \midrule
     Albrecht et al. \cite{10.1007/978-3-030-75539-3_16} & & \XSolidBrush & \XSolidBrush & \XSolidBrush & \XSolidBrush & & & & & &\\
     \rowcolor{lightgray!50} Ansong et al. \cite{Ansong} & & & \XSolidBrush & & & & & & & &\\
     Bassolé et al. \cite{Bassole} & \XSolidBrush & & \XSolidBrush & & & & & & \XSolidBrush \\
     \rowcolor{lightgray!50} Castle et al. \cite{Castle} & \XSolidBrush & & \XSolidBrush & \XSolidBrush & \XSolidBrush & & \XSolidBrush & & & &\\
     Chiboora et al. \cite{chiboora2023evaluating} & \XSolidBrush & \XSolidBrush & \XSolidBrush & & \XSolidBrush & & & & & \XSolidBrush & \XSolidBrush \\
     \rowcolor{lightgray!50} Ibrar et al. \cite{Ibrar} & \XSolidBrush & \XSolidBrush & & & \XSolidBrush & \XSolidBrush & & \XSolidBrush & & &\\
     Kaka et al. \cite{7947811} & & & & & \XSolidBrush & & & & \\
     \rowcolor{lightgray!50} Kant Kamal et al. \cite{10125814} & \XSolidBrush & \XSolidBrush & & & \XSolidBrush & & & & & & \XSolidBrush\\
     Koala et al. \cite{Koala} & & & \XSolidBrush & & & & & & \\
     \rowcolor{lightgray!50} Kumar et al. \cite{247668} & & & \XSolidBrush & \XSolidBrush & & & & & & &\\
     Latifa et al. \cite{10.1145/3108421.3108433} & & & \XSolidBrush & & & & & & & &\\
     \rowcolor{lightgray!50} Madwanna et al. \cite{9478078} & & & & & & & & & \XSolidBrush & & \\
     Munyendo et al. \cite{Munyendo} & & & \XSolidBrush & & & & & & & &\\
     \rowcolor{lightgray!50} Osho et al. \cite{Osho} & \XSolidBrush & & & & & & & & & & \\
     Reave et al. \cite{Reaves} & \XSolidBrush & \XSolidBrush & & \XSolidBrush & \XSolidBrush & & & & \\
     \rowcolor{lightgray!50} Shezan et al. \cite{7885802} & \XSolidBrush & & & & & \XSolidBrush & \XSolidBrush & \XSolidBrush & & \XSolidBrush & \\
     Sihag et al. \cite{10.1007/978-981-15-6648-6_20} & & & & & & & & & \XSolidBrush \\
     \rowcolor{lightgray!50} Uduimoh et al. \cite{Uduimoh} & \XSolidBrush & & & & & & & & & &\\
     Vakare et al. \cite{10060666} & \XSolidBrush & \XSolidBrush & & & \XSolidBrush & & & & \\
     \midrule
     \textbf{Number of publication} & \textbf{10} & \textbf{6} & \textbf{9} & \textbf{4} & \textbf{8} & \textbf{2} & \textbf{2} & \textbf{2} & \textbf{3} & \textbf{2} & \textbf{2}\\
  \bottomrule
\end{tabular}
\end{adjustbox}
\end{table}
This section enumerates the key security concerns of mobile apps in developing countries addressed in the literature.
As illustrated in table~\ref{tab:rq2}, the most common issues addressed relate to data storage, permissions, network communication, and cryptography, respectively 37\%, 29\%, 29\%, and 22\% of the publications. 
However, the literature has also identified security concerns regarding access control, inter-process communication (IPC), malware, WebView, third-party libraries, debug mode enabled, and root detection issues.

This information underscores the importance of addressing these concerns to mitigate the associated risks, such as privacy data leaks, unauthorized access, remote command execution, data interception, insecure data protection, etc.
As illustrated in Table~\ref{tab:rq2}, most of the publications have addressed security issues related to data storage, permissions, network communication, and cryptography.

Several studies reveal that apps insecurely store sensitive data in memory~\cite{Uduimoh}, log sensitive data clear text or with poor encryption~\cite{Castle, Reaves}, or allow data backup and enable other apps to read log files~\cite{Ibrar}.
This issue can easily compromise users' security since tools, such as ADB, can be used, for example, to do complete backups of the apps containing sensitive data~\cite{Argudo}.

Permissions in mobile apps help users to protect restricted access such as users' information, system state, audio/camera record, and connecting to a paired device~\cite{Android001}. However, one could use them to collect users' sensitive data~\cite{Munyendo}. 
Studies reveal that various apps request permissions to access resources, such as for writing and reading data in the external storage~\cite{Castle, Koala}, and abuse permissions, which most are categorized as risky~\cite{Bassole, Ansong}. 

Android uses the SSL/TLS protocol to secure communication by protecting the app's data in the network.
Several trusted Certificate Authorities in Android provide certificates for many servers.
In communication between a server, the app should verify the trustworthiness of the server by checking the certificate~\cite{Android002}. Unfortunately, this is not the case in some apps.
Indeed, studies reveal that several apps do not implement or properly use SSL/TLS certification validation~\cite{Ibrar, Reaves}. Occasionally, apps use insecure protocols, such as HTTP, for certain communication~\cite{Castle, Reaves}.
Even if this communication does not transport sensitive data, it can be used by attackers for man in a man-in-the-middle attack or by a phishing attack.

Studies reveal that cryptography is poorly used in several mobile apps.
Some apps use custom cryptography or poorly implemented cryptographic mechanisms~\cite{Reaves}. While others use insecure cipher modes such as ECB, outdated cryptographic protocols like MD5, SHA1, and DES, and hard-coded cryptographic keys to encrypt data~\cite{Ibrar, chiboora2023evaluating}.

Beside that, some concerns are not commonly addressed in the literature.
For example, there are security concerns related to code quality and resilience.
Code quality addresses coding vulnerabilities originating from external sources, including app data entry points, the OS, and third-party software components~\cite{masvs-cqb}. For instance, Android debug mode enables the transfer of data from the app to a computer and vice versa, as well as installing and uninstalling apps on the device, reading logcat memory, etc.
The developer can enable this mode on the app for testing purposes. However, if it is not disabled before its deployment, the app could allow attackers to inject code for malicious activities, access sensitive data, etc. In some studies, researchers found that this mode is enabled in several apps~\cite{7885802, chiboora2023evaluating}.
The resilience aspect covers vulnerabilities leading to app reverse-engineering, tampering, device jailbreak, etc. It verifies that the app operates on a trusted platform, prevents runtime tampering, and maintains the integrity of the intended app functionality~\cite{masvs-resilience}. Some authors have analyzed apps for resilience and found that several apps try to have root access on the device, do not implement any obfuscation methods, and do not prevent reverse-engineering~\cite{10125814, chiboora2023evaluating}.

These concerns affect mobile apps worldwide.
However, some studies addressed them by focusing on security challenges in developing countries~\cite{Castle}. 
\\

\begin{adjustbox}{width=\linewidth}
    \fcolorbox{black}{lightgray!50}{
        \begin{varwidth} {\linewidth}
            \textbf{RQ2:} In most publications, investigations are related to data storage, permissions, network communication, and cryptography, concerns that are not specific to developing countries.
            A few studies, however, have addressed various security concerns by focusing on specific challenges in developing countries.
        \end{varwidth}
    }
\end{adjustbox}

\subsection{Categories of apps}
\begin{table}[htb!]
    \caption{The categories of apps addressed in the literature.}
    \label{tab:rq3}
    \begin{adjustbox}{width=.9\linewidth,center}
        \begin{tabular}{l||cccccccccccc}
            \toprule
                Publication&Finance&Health\&Fitness&Education&Communication&Travel\&Local&Medical&Generic\\
                \midrule
                 Albrecht et al. \cite{10.1007/978-3-030-75539-3_16} & & & & \XSolidBrush & & & \\
                 \rowcolor{lightgray!50} Al-Ameen et al. \cite{10.1145/3378393.3402244} & & & & & & & \XSolidBrush \\
                 Ansong et al. \cite{Ansong} &\XSolidBrush & & & & & & \\
                 \rowcolor{lightgray!50} Bandan et al. \cite{9984416} & \XSolidBrush & & \XSolidBrush & \XSolidBrush & \XSolidBrush & \XSolidBrush & \\
                 Bassolé et al. \cite{Bassole} & \XSolidBrush & & & & & & \\
                 \rowcolor{lightgray!50} Castle et al. \cite{Castle} & \XSolidBrush & & & & & &\\
                 Chiboora et al. \cite{chiboora2023evaluating} & \XSolidBrush & & & & & &\\
                 \rowcolor{lightgray!50} Chopdar et al. \cite{CHOPDAR2022100651} & & \XSolidBrush & & & & & \\
                 Ibrar et al. \cite{Ibrar} & \XSolidBrush & & & & & & \\
                 \rowcolor{lightgray!50} Kaka et al. \cite{7947811} & \XSolidBrush & & & & & & \\
                 Kamdjoug et al. \cite{Kamdjoug} & \XSolidBrush & & & & & & \\
                 \rowcolor{lightgray!50} Kant Kamal et al. \cite{10125814} & & & & & & & \XSolidBrush \\
                 Khatoon et al. \cite{9012487} & & \XSolidBrush & & & & & \\
                 \rowcolor{lightgray!50} Koala et al. \cite{Koala} & & & & & & & \XSolidBrush \\
                 Kumar et al. \cite{247668} & \XSolidBrush & & & & & & \\
                 \rowcolor{lightgray!50} Latifa et al. \cite{10.1145/3108421.3108433} & \XSolidBrush & & & & & & \\
                 Madwanna et al. \cite{9478078} & \XSolidBrush & & & & & & \\
                 \rowcolor{lightgray!50} Munyendo et al. \cite{Munyendo} & \XSolidBrush & & & & & & \\
                 Osho et al. \cite{Osho} & \XSolidBrush & & & & & & \\
                 \rowcolor{lightgray!50} Prakash et al. \cite{PRAKASH2021100573} & & \XSolidBrush & & & & & \\
                 Reave et al. \cite{Reaves} & \XSolidBrush & & & & & & \\
                 \rowcolor{lightgray!50} Shezan et al. \cite{7885802} & & & & & & & \XSolidBrush \\
                 Sihag et al. \cite{10.1007/978-981-15-6648-6_20} & & & & & & & \XSolidBrush \\
                 \rowcolor{lightgray!50} Uduimoh et al. \cite{Uduimoh} & \XSolidBrush & & & & & & \\
                 Vakare et al. \cite{10060666} & \XSolidBrush & & & & & & \\
                 \midrule
                 \textbf{Number of publication} & \textbf{16} & \textbf{3} & \textbf{1} & \textbf{2} & \textbf{1} & \textbf{1} & \textbf{5} \\
            \bottomrule
        \end{tabular}
    \end{adjustbox}
\end{table}
As shown in Table~\ref{tab:rq3}, the SLR illustrates the primary interest in financial applications, particularly mobile banking and money apps.
Financial apps receive substantial attention from researchers. 
Some researchers analyzed the security vulnerabilities of mobile banking and payment apps within African countries aiming to create awareness among users, businesses, and governments about potential security threats~\cite{Bassole}, to assess how these apps handle sensitive user information, and to identify potentially exploitable vulnerabilities~\cite{Osho}. Others provide a comprehensive security analysis of the Unified Payments Interface used by some mobile payment apps in India, aiming to identify vulnerabilities and propose a more secure UPI for payment apps~\cite{9478078, 247668}. This can be motivated by the fact that they are relatively more sensitive to security compared to other categories of non-financial applications in developing countries~\cite{Bassole}, and studies talk about the numerous security challenges they face~\cite{approov2023} and the types of frauds related to mobile financial services in these regions~\cite{fouad2017}. 
Mobile apps related to Health\&Fitness receive the attention of researchers in the context of the COVID-19 pandemic through interviews and surveys to determine the factors that influence the adoption and the continuance intentions of usage of contact tracing apps~\cite{CHOPDAR2022100651, PRAKASH2021100573}. 
Aiming to provide secure access to health data and facilitate efficient interaction between patients and healthcare providers, some authors implement solutions integrating OAuth 2.0 and the Aadhaar identification system into Indian e-health apps~\cite{9012487}. 
Communicational apps are studied less frequently, as well as Educational, Travel \& Local, and Medical apps.
Several studies do not mention the category of apps they focused on. We have grouped those apps in the Generic category.
\\

\begin{adjustbox}{width=\linewidth}
    \fcolorbox{black}{lightgray!50}{
        \begin{varwidth} {\linewidth}
           \textbf{RQ3:} In the developing country context, researchers primarily focus on financial mobile apps, such as mobile banking and mobile money apps.
        \end{varwidth}
    }
\end{adjustbox}
\subsection{Security issue detection techniques}
\begin{table}[h]
    \centering
    \caption{Summary of the techniques used to detect security issues.}
    \label{tab:rq4}
    \begin{adjustbox}{width=0.65\linewidth,center}
        \begin{tabular}{l||cccc}
            \toprule
            Publication&Static&Dynamic&Hybrid&Forensic\\
            \midrule
             Albrecht et al. \cite{10.1007/978-3-030-75539-3_16} & & \XSolidBrush & & \\
             \rowcolor{lightgray!50} Ansong et al. \cite{Ansong} &\XSolidBrush & & & \\
             Bassolé et al. \cite{Bassole} & & & \XSolidBrush & \\
             \rowcolor{lightgray!50} Castle et al. \cite{Castle} & \XSolidBrush & & & \\
             Chiboora et al. \cite{chiboora2023evaluating} & & & \XSolidBrush & \\
             \rowcolor{lightgray!50} Ibrar et al. \cite{Ibrar} & \XSolidBrush & & & \\
             Kaka et al. \cite{7947811} & \XSolidBrush & & & \\
             \rowcolor{lightgray!50} Kant Kamal et al. \cite{10125814} & & & \XSolidBrush & \\
             Koala et al. \cite{Koala} & \XSolidBrush & & & \\
             \rowcolor{lightgray!50} Kumar et al. \cite{247668} & & & \XSolidBrush & \\
             Latifa et al. \cite{10.1145/3108421.3108433} & \XSolidBrush & & & \\
             \rowcolor{lightgray!50} Madwanna et al. \cite{9478078} & & & \XSolidBrush & \\
             Osho et al. \cite{Osho} & & & & \XSolidBrush \\
             \rowcolor{lightgray!50} Reave et al. \cite{Reaves} & \XSolidBrush & & & \\
             Shezan et al. \cite{7885802} & & & \XSolidBrush & \\
             \rowcolor{lightgray!50} Sihag et al. \cite{10.1007/978-981-15-6648-6_20} & & \XSolidBrush & & \\
             Uduimoh et al. \cite{Uduimoh} & & & & \XSolidBrush \\
             \rowcolor{lightgray!50} Vakare et al. \cite{10060666} & & & \XSolidBrush & \\
             \midrule
             \textbf{Number of publication} & \textbf{7} & \textbf{2} & \textbf{7} & \textbf{2} \\
            \bottomrule
        \end{tabular}
    \end{adjustbox}
\end{table}
This section explores the techniques employed to detect security concerns on mobile apps in developing countries.
Numerous methods exist for identifying vulnerabilities in these apps, with static and dynamic analysis being the most prevalent.
Static analysis involves examining software without its execution, while dynamic analysis involves the examination of software during runtime. 
These techniques are applied based on specific needs, and to enhance analysis efficiency, researchers frequently integrate static and dynamic analysis to create a hybrid approach.

The literature indicates a predominant reliance on static and hybrid analysis, as illustrated in table~\ref{tab:rq4}.

Additionally, dynamic analysis is occasionally utilized, along with forensic analysis, which entails investigating an app's traces in the device's memory.

The results show that most publications have only used off-the-shelf techniques and tools to uncover security issues~\cite{Ibrar, Bassole, Osho, 10125814, chiboora2023evaluating}. 
Others have combined those tools with specialized techniques to look for features leading to data leakage~\cite{Castle}, to analyze the communication flows between server and mobile apps~\cite{Castle, Reaves}, and the registration and transaction processes~\cite{9478078,247668}.
\\

\begin{adjustbox}{width=\linewidth}
    \fcolorbox{black}{lightgray!50}{
        \begin{varwidth} {\linewidth}
           \textbf{RQ4:} Researchers predominantly have employed static and hybrid analysis methodologies to scrutinize mobile apps for potential security issues. Nevertheless, most did not specifically adapt their works to specific concerns of developing countries.
        \end{varwidth}
    }
\end{adjustbox}

\subsection{Motivations for research}
With Smartphones becoming increasingly popular in developing countries\footnote{Source: \url{http://www.osiris.sn/En-Afrique-subsaharienne-le-taux-d.html}}, a surge in mobile app usage has raised awareness about security issues among researchers. 
Researchers motivate their studies by mentioning the growing concerns regarding mobile app security that impact many users in developing countries. 
They aim to encourage practitioners to enhance user protection and security measures, especially in the case of mobile financial services. 
The dearth of studies on security issues in developing countries inspires researchers to fill this gap. 
Furthermore, examining the relatively low adoption rates of certain mobile apps also fuels researchers' curiosity.
This curiosity prompted investigations into whether mobile app security plays a role in influencing adoption patterns.

In certain instances, researchers strive to reassure users, mitigate the threat posed by malicious apps, and address the unique security challenges prevalent in their respective regions. 
This multifaceted approach reflects a commitment to understanding, confronting, and ultimately improving the landscape of mobile app security in the context of developing countries.
\\

\begin{adjustbox}{width=\linewidth}
    \fcolorbox{black}{lightgray!50}{
        \begin{varwidth} {\linewidth}
           \textbf{RQ5:} Researchers generally motivate their papers by the increased adoption of smartphones coupled with a significant increase in mobile app usage, the growing concerns surrounding mobile apps, and the lack of studies in developing countries.
        \end{varwidth}
    }
\end{adjustbox}

%% file: discussion.tex
\section{Discussion}
\label{discuss}

This section discusses various aspects of mobile app security in developing countries.
We address unexplored research directions, emerging trends, and issues related to vulnerabilities and outline future challenges.

\subsection{Researches performed and needs}
In the domain of mobile app security in developing countries, several uncharted research directions hold the potential to drive innovation and enhance security practices.
Certain avenues have been much more explored. 
\\

\textbf{Explored Research directions}
\\
The identified papers investigate a few relevant research avenues that could help develop more secure apps.  

Several research papers have conducted user studies through interviews and surveys, which offer useful insights into user behavior and perceptions when using mobile apps, and their intentions to use certain mobile apps. This type of study provides a deep understanding of user experiences and sentiments.
 It also guides the development of more secure apps since it has allowed researchers to provide actionable recommendations and guidelines for developers and policymakers, as well as app markets.
 Researchers use advanced methods and frameworks such as machine learning algorithms, UTAUT, and PLS-SEM to provide robust analysis. However, we observed that they often use a limited sample size. The sample size used could be representative, taking into account gender diversity and geographic area, to provide stronger results. We have noticed that researchers primarily explore the adoption of financial and healthcare apps. This is understandable since they are two crucial domains. Nevertheless, the scope could be extended, allowing it to cover other domains.

Researchers have mainly focused on app analysis approaches, such as static, dynamic, hybrid, and forensic analysis.
Their work is primarily based on using off-the-shelf automated security detection tools, among which some have limited capabilities to detect specific vulnerabilities~\cite{Ibrar}.
Each study's focus is on specific security aspects according to the category of apps it addresses.
We have noticed that financial apps have received more attention from researchers than other categories of apps.
This attention is not driven by the fact that financial apps are the most used in developing countries, so researchers should attach importance to mobile apps in other domains, including healthcare, gaming, and social media, since the security of these apps could affect millions of users.

Researchers have also minimally addressed app security testing~\cite{10125814, 10060666}.
This type of study is useful since it not only allows the identification of vulnerabilities statically but also provides a runtime analysis of vulnerabilities, raising awareness in users as well as developers side.
According to the threats targeting mobile apps in developing countries~\cite{statica001, approov2023, fouad2017}, there is a pressing need for specialized solutions to deal with these problems. Unfortunately, only a single study has been provided proposing a context-specific tool to ensure secure access, authentication, and authorization within the healthcare domain in India~\cite{9012487}. Another work has proposed a system that collects and evaluates behavioral activity for malicious intent from targeted apps~\cite{10.1007/978-981-15-6648-6_20}.
However, these approaches are not specific to developing countries. 
The existing literature has not mentioned any security testing tools specifically tailored for mobile apps in developing countries.
Researchers typically use pre-existing open-source tools developed for a global context, even for app analysis, rather than creating region-specific solutions.

We have noticed a significant lack of research into the security of mobile applications in developing countries.
More studies should be proposed, ranging from considering more app categories to proposing solutions adapted to developing countries.
Many approaches exist in the literature worldwide. 
However, applying them in the developing country context can lead to miss interesting findings because the types of functionality and errors in mobile apps could be different.
\\

\textbf{Unexplored Research directions}
\\
Additionally, there is a domain that deserves to be more explored. 
Mobile malware detection should be explored in greater depth. Indeed, mobile phone users across the globe are increasingly falling victim to a rising tide of mobile malware infections~\cite{baur2019cyber}. 
These infections disproportionately impact users in developing countries. 
According to Carter~\cite{carter2017forces}, the top ten countries most severely affected by mobile malware are developing countries, meaning that it deserves important attention from researchers.
Despite this, we have found only two studies focusing on mobile malware detection in these regions. The one designed a system that analyzes real-time app interactions for malicious behavior~\cite{10.1007/978-981-15-6648-6_20}, while the other one conducted a vulnerability analysis using VirusTotal for malware detection~\cite{Bassole}. 
While several advanced malware detections exist, using AI techniques such as machine learning and deep learning, these studies primarily leveraged dynamic analysis techniques to address mobile malware~\cite{10.1007/978-981-15-6648-6_20}.
Apart from these two works, no other studies have delved into the critical issue of mobile malware detection.
It is noteworthy that researchers tend to focus on analyzing mobile apps for security vulnerabilities rather than proactively addressing the issue of mobile malware detection despite the escalating threat posed by mobile malware in developing countries.
To better point out the gap in this research direction, a study could be performed, exploring the types of malware used, the payload of these malware, the techniques and procedures used by attackers to compromise devices in these regions, the active threat actors, and the information assets that are attacked the most as well. This may help researchers to understand the contributions needed to address the gap.

Another research direction could be to explore proofs-of-concept allowing the exploitation of vulnerabilities found in mobile apps. This can help to improve its security. 
In our study, we have identified only one paper in which authors have outlined hypothetical attacks on mobile money apps without demonstrating how these attacks could be executed~\cite{Castle}.
This hesitancy may be attributed to the constraints inherent in conducting proof-of-concept experiments. In fact, various factors, such as governmental regulations, resource constraints, or the requisite access to specific accounts, can hinder the practical exploitation of vulnerabilities. However, presenting straightforward use cases can potentially address these issues. Despite these challenges, investigating how adversaries can exploit vulnerabilities remains crucial for improving mobile app security in developing countries.

\subsection{Research trends}
We were surprised by the few papers we have identified in this SLR.
Indeed, we have identified twenty-five publications, and the first appeared in 2016~\cite{Castle}.
In general, there is an average of three relevant publications annually until 2023, indicating the scarcity of research in this area.
To resolve this problem, more comprehensive work is needed to address the numerous security issues in the ever-growing mobile app landscape in developing countries.
In Section~\ref{chalenge}, we proposed several future types of research identified when analyzing publications to help researchers fill the gap.

\subsection{Future challenges}
\label{chalenge}
Our discussion highlights perspectives from the existing literature and presents new challenges that need to be addressed to enhance mobile app security in developing countries.
\\

\textbf{Challenges from the literature.}
\\
The authors of the papers listed in our SLR have identified several open challenges to extend research or explore new directions. The identified perspectives include:
\begin{enumerate}
    \item \textit{Investigating development frameworks.} Authors explored some app development frameworks to understand the impacts they have on the security of the apps. They argue that certain frameworks allow the development of more secure apps than others. The study focuses on two development frameworks, and the authors suggested extending it by considering more frameworks.
    \item \textit{Reducing the requested permissions}. Some studies investigated mobile app security by focusing on permissions. The authors found that some apps abuse users by requesting unnecessary permissions, which could have an impact on their security. These authors suggest investigating and proposing solutions to reducing the permissions requested by apps to protect user privacy.
    \item \textit{Developing alert tools.} There are also perspectives to develop tools to alert users and administrators about potentially dangerous permissions used by apps.
    \item \textit{Expanding studies.} When performing interviews and surveys, researchers used a small number of participants and focused on specific areas and institutions. To have more comprehensive results, they suggested expanding the studies by covering diverse contexts, institutions, and participants.
    \item \textit{Focusing on specific aspects.} To refine the analysis results, researchers suggested carrying out studies that focus on specific aspects of financial apps, such as password reset procedures and accountless money transfers.
\end{enumerate}

\textbf{Future research directions.} 
\\
Based on the insufficiently explored research directions mentioned above and the need for more specific approaches, we have identified, in addition to the challenges in the literature, the following future research directions:
\begin{enumerate}
    \item Investigate unexplored research areas on mobile app security analysis targeting developing countries' specific challenges.
    \item Extend applications of state-of-the-art research results to explore security concerns beyond fintech apps to other critical categories, such as health and education apps.
    \item Investigate malware targeting developing countries. In particular, analysis techniques should be developed that are suited for (1) pre-installed apps in low-cost devices and (2) available and still-in-use insecure technologies (e.g., USSD).
\end{enumerate}

Mobile app security in developing countries is an evolving field with numerous opportunities for research and improvement. Addressing these research directions and challenges is essential to enhance the security of mobile apps and protect users in these regions.

%% file: threat_to_validity.tex
\section{Threat to Validity}
\label{validity}

Our primary objective was to assemble publications discussing mobile app security in developing countries. 
However, there are potential threats to the validity of this study.

We established exclusion criteria to eliminate irrelevant publications during the selection process. Notably, we excluded books and thesis reports, as we deemed them less relevant, assuming that pertinent information would likely be found in conference or journal papers.

To streamline our search on the Springer digital library, we applied filters to select publications within the field of Computer Science. Subsequently, we filtered by conference papers and articles, focusing solely on these publications. Additionally, we excluded publications that were not written in English. While these actions were motivated by a desire to refine our results, they may have inadvertently excluded some potentially relevant publications.

We modified our search keywords in an effort to capture more relevant publications compared to our initial attempt. However, this adjustment could have resulted in the omission of some pertinent publications.

During the selection process, we excluded publications primarily centered on security infrastructure or systems in which mobile app technology was just a component. For instance, publications addressing mobile money infrastructure or learning systems were excluded, as they did not primarily focus on mobile apps. While this was done to maintain the focus of our study, it may have resulted in the unintentional exclusion of publications with relevant insights into mobile app security in the broader context.

Similarly, when assessing the quality of publications, we have excluded some publications.
This exclusion is not because the publications did not investigate mobile app security in developing countries.
We excluded them because they did not meet our quality criteria, potentially including biases in our dataset.
We did not exclude the possibility of having biases since we are only based on criteria we established.
Thus, we can miss some publications that deserve to be excluded or kept.
Even when analyzing data, we did not see additional publications to exclude since each enables us to answer at least one of our research questions.

%% file: related_work.tex
\section{Related Works}
\label{related}

To our knowledge, no systematic literature review focuses on mobile app security in developing countries. However, several reviews, investigations, and surveys have been performed on other topics in the context of developing countries.

Hsu et al.~\cite{hsu2016top} gave an overview of Chinese mobile health (mHealth) apps in December 2015 by investigating the most downloaded apps from Android and iOS. In their study, authors aimed to understand the current state of the Chinese mHealth market, focusing on medical-related apps categorized by ten different medical initiatives rather than general health apps.
For each app, they analyzed the main service offered, mHealth initiative, disease and specialty focus, app cost, target user, Web app availability, and emphasis on information security.
The findings revealed that the primary mHealth initiatives targeted by the apps reflect Chinese patients’ demand for access to medical care. Disease-specific apps are also representative of disease prevalence in China, with the most common disease-specific apps focusing primarily on diabetes, hypertension, and hepatitis management. Most apps were found to be free and available on both iOS and Android platforms. 
The paper also underlines the lack of information about users’ data security since developers did not mention the purposes for which users’ data could be used. 

Latif et al.~\cite{latif2017mobile} presented a study that underlines the factors hindering the use of mobile health in developing countries by reviewing various mobile health initiatives, their impacts, and their healthcare challenges. The review highlights several challenges that hinder the successful deployment of mHealth, including infrastructural limitations, the need for strategic partnerships, and cultural and language issues.
The authors emphasized the importance of leveraging emerging technologies such as the Internet of Things (IoT) and artificial intelligence (AI) to enhance mHealth adoption.
Additionally, the paper features a case study on Pakistan to validate the findings and illustrate the practical implications of mHealth in a specific context.
This case study demonstrates that tailored approaches, considering local cultural contexts, are essential for effective mHealth implementation.

Abdullaev et al.~\cite{abdullaev2019security} performed a state-of-the-art survey of mobile banking. They aimed to classify and analyze mobile banking services' challenges and security issues in Uzbekistan. They explored the security issues related to the Wireless Application Protocol (WAP). With this protocol, customers can use bank services through the Internet. However, data is not well encrypted at one stage of the communication using WAP. They also identified other risks and issues related to the authentication process, SMS banking, and virus attacks, in which attackers can use many techniques to steal users’ sensitive information. They presented many security risks that can compromise the mobile banking system.
In their study, authors also found that 60\% of customers do not use mobile banking technology.

Azeez and Lakulu~\cite{azeez2019review} performed a literature review on mobile government (m-government) in developing countries focusing on identifying the benefits, challenges, and critical success factors for the successful implementation of m-government from both government and citizen perspectives.
The authors analyzed various studies on the success of m-government and categorized the success factors based on their impact on the successful implementation of m-government. 
The findings reveal increased efficiency of governmental activities, cost reduction in organizations, and accessibility of government services as benefits of m-government, as well as the offer of real-time information and improvement of citizen participation in governmental activities.
However, the study identified several m-government challenges, including security and privacy, technical limitations, infrastructure limitations, and interoperability.
The paper concluded that while m-government offers significant opportunities for improving government services in developing countries, it also presents several challenges that need to be addressed to ensure successful implementation.

Hoque et al.~\cite{hoque2020mobile} present a review of studies related to mobile health applications between 2013 and 2018. Through this review, they aimed to evaluate the quality of evidence reporting by using mobile health Evidence Reporting and Assessment (mERA), a checklist developed by the World Health Organization (WHO). This review highlights the application level of evidence reporting as recommended by WHO. 
Authors found that researchers and mobile health intervention designers from developing countries have limited familiarity with the mobile health Evidence Reporting and Assessment checklist. 
They also noticed that the majority of studies fail to meet the essential evidence-reporting criteria outlined in the checklist. Furthermore, design science-based methods and theory-based frameworks are rarely applied in the development of mobile health interventions. Finally, they found that most mobile health interventions are not prepared for interoperability or integration into existing health information systems. Overall, the study reveals that most studies do not properly apply WHO's recommendations.

There is also another state-of-the-art in which Rahman et al.~\cite{rahman2020state} explore the current situation of mobile banking in Bangladesh. The research objectives are to identify the problems and challenges that customers face in mobile banking services and to observe the future prospects of mobile banking in the country.
The paper identified several problems related to security (financial loss across virus attacks, fraudulent activities, privacy leakage, etc.), time (because of late payments or any other reason, there is time lost), network (poor network), performance (server break down), financial (lose of money because of mistakes made during money transaction), and lack of banking knowledge. 
Authors also argued that issues may arise because of compatibilities (difficulty to operate with another device when the first breaks down, unable to access service in some places because of poor internet connection). 
According to them, trusting mobile banking is an important factor that may give issues such as fairness, capability, and beneficence.Despite these challenges, the paper highlighted the prospects of mobile banking, including benefits for phone operator companies, increasing job scopes, no service charge, increased purchasing power, and easy money transfer. The authors concluded that if these issues can be solved in the future, mobile banking will lead to a greater impact on the banking economy of Bangladesh.

Msweli et al.~\cite{msweli2020enablers} provided a study that explores the enablers and barriers to mobile banking and mobile commerce among the elderly, particularly in developing countries, by reviewing the existing literature focusing on literature from 2009 to 2019. The findings reveal that there are significant gaps in research concerning the elderly and mobile banking, particularly in developing countries. Key barriers identified include security concerns, lack of trust, and limited technical knowledge among older adults, which hinder their adoption of mobile banking services. Conversely, enablers such as the perceived ease of use and the potential for improved quality of life through mobile commerce were noted. The study concludes that there is a pressing need for further research to address these gaps and to develop tailored mobile banking solutions specifically to the needs of the elderly, thereby enhancing their financial inclusion in the digital era.
Similarly, Pankomera and van Greunen~\cite{Pankomera} systematically reviewed the opportunities, barriers, and adoption factors of mobile commerce (m-commerce) services for the informal sector in Africa. This paper seeks to provide comprehensive insights into how m-commerce can benefit the informal sector, the challenges faced, and the factors influencing its adoption.
The study identified several barriers, including the limited network coverage and broadband infrastructure, high costs of mobile devices and services, resistance due to illiteracy, lack of trust and traditional business practices, and finally, a lack of legal and regulatory frameworks to support m-commerce.
The adoption factors identified from the literature include network coverage and availability of electricity, knowledge of m-commerce and its benefits, perceived usefulness and ease of use, affordability of m-commerce solutions, confidence in the security of m-commerce transactions, accessibility of financial services, the impact of social and cultural factors on adoption, and government policies and regulations promoting m-commerce.
Furthermore, the authors identified several benefits and opportunities of m-commerce, such as the creation of new services, increased revenue, reduced operational costs, and enhanced productivity and market access, particularly in the agriculture and fishing sectors.

Malik~\cite{malik2020review} performs a review article in which he studies publications that talk about Internet and mobile banking adoption from 2015 to 2020 using the Unified Theory of Acceptance and Use of Technology (UTAUT) model in developing countries. This study highlights the directions, the most used analysis tools, and the main indicators of behavioral intention used in the publications. It also reveals that most publications focused on factors affecting Internet banking adoption (54\%). This study argues that the extended UTAUT model is mostly used in publications. Indeed, as presented, this model is used by 67\% of the publications among the 54\% internet banking and by 70\% among the 46\% mobile banking publications. This is similar to the study that reviews the literature regarding mobile health adoption in developing countries~\cite{aljohani2021adoption}. In this study, Aljohani et al. identify the methodologies used in existing research, the significant factors influencing adoption, and the gaps in the literature, particularly regarding the use of qualitative and mixed methods. 
The authors conducted a systematic review of the literature by evaluating various studies published between 2010 and 2020.
The findings revealed a limited number of studies on m-health adoption in developing countries, with a notable concentration of research in China and Bangladesh.
The Technology Acceptance Model (TAM) was frequently used, focusing primarily on technological and individual factors, while other health-related factors and theories were underexplored. Furthermore, most studies employed quantitative methodologies, with only one study utilizing a qualitative approach and none using mixed methods.
The authors emphasized the need for more qualitative and mixed-method studies to provide richer insights into the adoption of m-health applications and to better understand the cultural and health impacts of these technologies.

All of these mentioned studies are completely different from ours because we have mainly focused on investigating studies that have been conducted in this area.

%% file: conclusion.tex
\section{Conclusion}
\label{concl}

This SLR has delved into mobile app security in developing countries. We meticulously examined 25 publications from various conference and journal venues throughout this process. Our goal was to shed light on the diverse research directions, security concerns, categories of apps addressed, investigation techniques, and researcher motivations in the domain of mobile app security within developing countries.
Key findings from this SLR include:
\begin{enumerate}
    \item \textbf{Lack of studies}. The literature on mobile app security in developing countries is relatively scarce, highlighting a significant research gap in this crucial area.
    \item \textbf{Limited research directions}. The review identified two main research types: user study and app analysis. However, other promising directions, such as malware detection, development framework study, and app security testing, remain underexplored.
    \item \textbf{Focus on financial apps}. The literature mainly focuses on mobile apps related to finance, with fewer publications delving into the health and education sectors, indicating a potential avenue for future research diversification.
    \item \textbf{Lack of specific analyses}. Notably, there is a lack of sophisticated analyses specifically targeting mobile app security in developing countries, which could open new perspectives. 
\end{enumerate}

In addition to these insights, we have outlined some unaddressed future challenges that warrant further exploration in the field of mobile app security within developing countries.

It is essential to acknowledge that this study is limited to developing countries, excluding emerging ones. Future SLRs may expand on this work by including emerging countries, providing a more comprehensive perspective on mobile app security in diverse socio-economic contexts.

%% file: appendix.tex
\newpage
\appendix
\section{Appendix}
\label{app}

\begin{center}
\textbf{Summary of the selected publications}
\end{center}

\hrulefill
\newline

\noindent
\textbf{Paper Title \& Reference:} \textit{Effective Security Testing of Mobile Applications for Building Trust in the Digital World} \cite{10125814}

\textbf{Type of Study:} App Security Testing

\textbf{Summary:} The paper discusses the security testing procedures and features of India’s first indigenous AppStore, “mSeva AppStore”. It tests the security of 300 apps from this AppStore, for improper platform usage, insecure coding practices, insecure communication, and more, using several tools such as MobSF, Appium, Drozer, and Robotium. The test reveals common issues, including insecure communication, insufficient cryptography, and insecure data storage.

\textbf{Strengths:}
\begin{itemize}
\item Provides a detailed methodology for security testing.
\item Provides useful insights and results.
\item Offers a classification of common security issues.
\item Highlights the importance in ensuring mobile app security.
\end{itemize}

\textbf{Limitations:}
\begin{itemize}
\item Limit scope (AppStore and OS).
\item Sample size is not broad.
\end{itemize}

\hrulefill
\newline

\noindent
\textbf{Paper Title \& Reference:} \textit{User’s Perception on Security and Privacy in Using Crypto Currency
Trading Application in India} \cite{10060666}

\textbf{Type of Study:} App Security Testing and User study

\textbf{Summary:} The paper investigates the security and privacy perceptions of users regarding cryptocurrency trading apps in India. It focuses on the top 6 cryptocurrency trading apps and examines their security profiles using various penetration testing techniques. In this study, authors use supervised machine learning algorithms like Random Forest and Logistic Regression to analyze user reviews and sentiments, and the OWASP mobile security testing framework to conduct a security analysis on the apps to identify common
vulnerabilities. The study identifies several vulnerabilities related to data stoarge, cryptography, and network communication. Authors' sentiment analysis reveals a mix of positive and negative sentiments towards the security and privacy of these apps. Users’ perceptions of security and privacy are influenced by their awareness and knowledge of potential threats.

\textbf{Strengths:}
\begin{itemize}
\item Provides a comprehensive analysis.
\item  Uses machine learning algorithms to analyze user sentiments.
\item The findings have practical implications for developers and users.
\item Highlights areas for improvement in app security and user education.
\item Provide a clear understanding of the security issues in these apps.
\end{itemize}

\textbf{Limitations:}
\begin{itemize}
\item App sample size may not be representative.
\item Limit scope (OS and geography).
\item  Dependence on user reviews from app stores.
\end{itemize}

\hrulefill
\newline

\noindent
\textbf{Paper Title \& Reference:} \textit{Desperate
Times Call for Desperate Measures”: User Concerns with Mobile Loan
Apps in Kenya}~\cite{Munyendo}

\textbf{Type of Study:} User study

\textbf{Summary:} The paper investigates the usage, privacy concerns, and user behavior associated with mobile loan applications in Kenya. These apps, which offer quick and easy access to small loans, often at high interest rates, collect sensitive user data through permissions. The study uses semi-structured interviews with 20 users to explore their concerns and trade-offs between privacy and the need for loans.

\textbf{Strengths:}
\begin{itemize}
\item Provides useful insights into user behavior and concerns.
\item Offers a deep understanding of user experiences and privacy concerns.
\item Provides recommendations for regulators, developers, and app markets to improve user privacy and security.
\end{itemize}

\textbf{Limitations:}
\begin{itemize}
\item Limit scope (App category and geography).
\item Does not use a representative sample.
\end{itemize}

\hrulefill
\newline

\noindent
\textbf{Paper Title \& Reference:} \textit{Adoption of Covid-19 contact tracing app by
extending UTAUT theory: Perceived disease threat as moderato}~\cite{CHOPDAR2022100651}

\textbf{Type of Study:} User study

\textbf{Summary:} The study proposes a research model based on the Unified Theory of Acceptance and Use of Technology (UTAUT), Health Belief Model (HBM), perceived privacy risk, and perceived security risk to understand the adoption of contact tracing applications. An online survey was conducted among 307 users of the ‘Aarogya Setu’ app. The data was analyzed using Partial Least Squares Structural Equation Modelling (PLS-SEM). The study provides insights into both the drivers and barriers to the adoption of contact tracing apps. In particular, findings revealed that perceived privacy and security risks were significant barriers to the adoption.

\textbf{Strengths:}
\begin{itemize}
\item Integrates multiple theoretical frameworks (UTAUT, HBM) and perceived risks, providing a robust model.
\item The use of PLS-SEM provides strong empirical support for the proposed hypotheses.
\item Provides useful insights for app developers and policymakers.
\item Highlights the moderating role of perceived disease threat in the adoption of contact tracing apps.
\end{itemize}

\textbf{Limitations:}
\begin{itemize}
\item Limit scope (App category, geography, etc.).
\item Limits on the ability to capture changes in user behavior over time.
\item The sample size may not be representative.
\end{itemize}

\hrulefill
\newline

\noindent
\textbf{Paper Title \& Reference:} \textit{State of Survey: Advancement of Knowledge Environmental Sustainability in Practicing Administrative Apps}~\cite{9984416}

\textbf{Type of Study:} User study

\textbf{Summary:} The paper presents a survey of government mobile apps in Bangladesh, focusing on user satisfaction and the factors contributing to the popularity or unpopularity of these apps. The survey was conducted with 310 participants to gather insights on their experiences and opinions regarding government apps. The findings indicate a general dissatisfaction among users, primarily due to issues related to security, content quality, design, and the presence of bugs.

\textbf{Strengths:}
\begin{itemize}
\item Focuses on user feedback.
\item Clearly identifies key areas of dissatisfaction.
\item Offers actionable suggestions for enhancing app features.
\end{itemize}

\textbf{Limitations:}
\begin{itemize}
\item The sample size may not be representative.
\item Limit scope ((App category and geography).
\item Potential bias in responses.
\item Does not account for changes in user perceptions or app performance over time.
\end{itemize}

\hrulefill
\newline

\noindent
\textbf{Paper Title \& Reference:} \textit{Understanding digital contact tracing app continuance: Insights from India}~\cite{PRAKASH2021100573}

\textbf{Type of Study:} User study

\textbf{Summary:} The paper investigates the factors influencing the continuance intentions of users regarding digital contact tracing (DCT) apps in India, particularly in the context of the COVID-19 pandemic. It extends the Expectation-Confirmation Model (ECM) by incorporating additional factors such as trust in technology and perceived security and privacy. The study employs a quantitative approach, utilizing a survey distributed to users of the "Aarogya Setu" app, and analyzes the data using Partial Least Squares Structural Equation Modeling (PLS-SEM). The findings reveal that user satisfaction, trust in the DCT app, and trust in the government are significant determinants of users' intentions to continue using the app.

\textbf{Strengths:}
\begin{itemize}
\item The use of PLS-SEM allows for a robust analysis of the relationships between constructs.
\item Adheres to established guidelines for assessing measurement and structural models.
\item Offers recommendations for policymakers and app developers to improve user engagement and trust.
\end{itemize}

\textbf{Limitations:}
\begin{itemize}
\item Sample bias (demography).
\item Does not account for changes in user intentions or behaviors over time.
\item Limit scope (geography).
\item Focus on intentions rather than actual usage behavior.
\end{itemize}

\hrulefill
\newline

\noindent
\textbf{Paper Title \& Reference:} \textit{Determining factors and impacts
of the intention to adopt mobile banking app in Cameroon: Case of
SARA by afriland First Bank}~\cite{Kamdjoug}

\textbf{Type of Study:} User study

\textbf{Summary:} The paper investigates the adoption and use of mobile banking apps, specifically focusing on the SARA app by Afriland First Bank in Cameroon. It employs various theoretical frameworks, including the Technology Acceptance Model (TAM), Protection Motivation Theory (PMT), and the Unified Theory of Acceptance and Use of Technology (UTAUT2), to understand the factors influencing users' decisions to adopt mobile banking technology. The study utilizes a structured questionnaire with a Likert scale to gather data from users, analyzing responses through Partial Least Squares Structural Equation Modeling (PLS-SEM).  findings indicate that factors such as perceived ease of use, perceived usefulness, and security concerns significantly impact users' intentions to adopt mobile banking apps.

\textbf{Strengths:}
\begin{itemize}
\item Integrates multiple established theories to create a comprehensive model.
\item Focuses on gender differences.
\item The findings help mobile banking service providers to meet user needs and enhance adoption rates.
\end{itemize}

\textbf{Limitations:}
\begin{itemize}
\item Limit scope (App and geography).
\item The sample size may not be representative.
\item Lack of qualitative exploratory research.
\end{itemize}

\hrulefill
\newline

\noindent
\textbf{Paper Title \& Reference:} \textit{We Don’t Give a Second Thought
Before Providing Our Information: Understanding Users’ Perceptions of
Information Collection by Apps in Urban Bangladesh}~\cite{10.1145/3378393.3402244}

\textbf{Type of Study:} User study

\textbf{Summary:} The paper investigates the perceptions of users in urban Bangladesh regarding data collection by smartphone apps. It highlights the varying attitudes towards privacy, ranging from indifference to fear, and examines how local infrastructure and social practices influence these perceptions. The study is based on interviews with 32 participants from diverse backgrounds in Dhaka, Bangladesh. The findings reveal that participants were aware of privacy issues but had mixed feelings about data collection. Some saw it as beneficial or necessary, while others felt indifferent or fearful. Apps are seen as essential for navigating city life, despite privacy concerns. Language barriers and the complexity of privacy policies hinder users’ understanding and ability to make informed decisions.

\textbf{Strengths:}
\begin{itemize}
\item Provides useful insights into privacy perceptions.
\item Captures a broad spectrum of perspectives by including participants from various age groups, literacy levels, and professions.
\item Proposes recommendations to improve privacy awareness and practices tailored to the local context of Bangladesh.
\item Integrates perspectives from human-computer interaction, security, and social sciences.
\end{itemize}

\textbf{Limitations:}
\begin{itemize}
\item A small sample of participants.
\item Limit scope (Area and language).
\item The reliance on self-reported data may introduce biases.
\end{itemize}

\hrulefill
\newline

\noindent
\textbf{Paper Title \& Reference:} \textit{Let’s Talk Money: Evaluating the Security Challenges
of Mobile Money in the Developing World}~\cite{Castle}

\textbf{Type of Study:} User study and App analysis

\textbf{Summary:} The paper investigates the security challenges of mobile money services in the developing world. includes a large-scale analysis of 197 Android apps and interviews with 7 developers from Africa and South America. The authors propose a systematic threat model to assess potential attacks and evaluate current security practices. The security analysis reveals that many apps have vulnerabilities, but service providers generally make security-conscious decisions.

\textbf{Strengths:}
\begin{itemize}
\item Provides a comprehensive analysis supported by both app analysis and qualitative insights from developer interviews.
\item Proposes a systematic threat model, helping in identifying potential attacks on DFS applications.
\item Offers recommendations to improve security practices, including the use of standard encryption and better training for developers.
\end{itemize}

\textbf{Limitations:}
\begin{itemize}
\item Limit scope (Android OS).
\item Sample size for interviews limited.
\item The insights from developer interviews rely on self-reported data, which may be subject to bias or inaccuracies.
\item Regulatory constraints.
\end{itemize}

\hrulefill
\newline

\noindent
\textbf{Paper Title \& Reference:} \textit{Evaluating Mobile Banking Application
Security Posture Using the OWASP’s MASVS Framework}~\cite{chiboora2023evaluating}

\textbf{Type of Study:} User study and App analysis

\textbf{Summary:} The paper presents an analysis of 18 mobile banking apps from various financial institutions in Africa. The study aims to assess the security posture of these apps using the OWASP Mobile App Security Verification Standard (MASVS) v2.0 framework. It combines manual and automated testing to analyze app source code and performs developer surveys. The paper identifies several security vulnerabilities in the tested apps, including issues with data storage, cryptography, network communication, permissions, code quality, and resilience. It also highlights the challenges developers face in implementing security measures, such as lack of expertise, time constraints, and complexity.

\textbf{Strengths:}
\begin{itemize}
\item Use of comprehensive framework (OWASP MASVS).
\item Analyzes apps from various financial institutions across different regions in Africa.
\item Provides a detailed breakdown of the security issues found in each category of the MASVS framework.
\item Highlights the practical challenges faced by developers in implementing security measures.
\end{itemize}

\textbf{Limitations:}
\begin{itemize}
\item Focuses solely on unauthenticated client-side testing.
\item A larger sample size could provide more generalizable results.
\item Limit scope (geography and app category).
\item Limited profiling of respondents for the surveys.
\end{itemize}

\hrulefill
\newline

\noindent
\textbf{Paper Title \& Reference:} \textit{Analysis of the Impact of Permissions on the
Vulnerability of Mobile Applications}~\cite{Koala}

\textbf{Type of Study:} App analysis

\textbf{Summary:} The paper investigates the security risks associated with permission management in Android applications, particularly focusing on apps developed in Burkina Faso. It aims to identify permission abuses and propose measures to enhance data protection for users, developers, and administrators. The study involves statically analyzing a sample of 40 apps developed in Burkina Faso, covering various categories available on Google Play. The findings reveal that despite existing security measures, many apps still exhibit vulnerabilities related to permission management. The paper emphasizes the need for developers to improve their UID allocation systems, limit the number of signatures per app, and ensure that apps with similar features use consistent permission groups.

\textbf{Strengths:}
\begin{itemize}
\item Analyzes a diverse sample of apps across various categories.
\item Identifies specific permission abuses and their potential impact on user data.
\item Provides actionable recommendations for developers and users' awareness to mitigate risks.
\end{itemize}

\textbf{Limitations:}
\begin{itemize}
\item Limit scope (geography).
\item Limited sample size.
\item Lack of User Perspective.
\end{itemize}

\hrulefill
\newline

\noindent
\textbf{Paper Title \& Reference:} \textit{A Study of Static Analysis Tools to Detect Vulnerabilities of Branchless Banking Applications in Developing Countries}~\cite{Ibrar}

\textbf{Type of Study:} App analysis

\textbf{Summary:} The paper investigates the effectiveness of static analysis tools in identifying security vulnerabilities in Android Digital Financial Services (DFS) apps. It focuses on apps used in developing countries and compares them with those from developed countries. The authors used 3 static analysis tools: MobSF, Qark, and AndroBugs, to analyze 10 DFS apps. They found higher vulnerability in DFS apps in developing countries than in developed countries and argued that off-the-shelf static analysis tools have limitations, especially for DFS-specific vulnerabilities. They identified common vulnerabilities such as insecure inter-process communication, insecure data storage, inadequate use of cryptography, insecure network communication, and WebView vulnerabilities.

\textbf{Strengths:}
\begin{itemize}
\item Provides a comprehensive analysis.
\item Provides comparative study by comparing apps from developing and developed countries.
\item Offers insights for developers and policymakers to enhance the security of DFS apps, particularly in developing countries.
\item Identifies specific categories of vulnerabilities, providing a clear focus for future improvements.
\end{itemize}

\textbf{Limitations:}
\begin{itemize}
\item Inherits tool limitations in detecting SSL/TLS misconfigurations and runtime issues.
\item The sample may not be representative.
\item Limit scope (Android OS).
\end{itemize}

\hrulefill
\newline

\noindent
\textbf{Paper Title \& Reference:} \textit{Vulnerability Analysis in Mobile Banking and Payment Applications on Android in African Countries}~\cite{Bassole}

\textbf{Type of Study:} App analysis

\textbf{Summary:} The paper assesses the security vulnerabilities of mobile banking and payment apps specifically on Android platforms within African nations, identifying risks related to privacy and data confidentiality. It aims to create awareness among users, businesses, and governments about potential security threats while encouraging the integration of robust security measures in the development of mobile apps.

\textbf{Strengths:}
\begin{itemize}
\item Provides a detailed examination of vulnerabilities, focusing on access permissions and code vulnerabilities.
\item Gives awareness to stakeholders about the risks.
\item Offers insights for developers and users to enhance security.
\end{itemize}

textbf{Limitations:}
\begin{itemize}
\item Limit scope (App and geography).
\end{itemize}

\hrulefill
\newline

\noindent
\textbf{Paper Title \& Reference:} \textit{Forensic Analysis of Mobile Banking Apps}~\cite{Osho}

\textbf{Type of Study:} App analysis

\textbf{Summary:} In this paper, the authors investigate the security and forensic aspects of mobile banking apps, particularly focusing on popular Android banking apps in Nigeria. The study aims to assess how these apps handle sensitive user data and to identify potential vulnerabilities that could be exploited by attackers. The authors conducted a thorough analysis of twelve mobile banking apps, examining their data storage practices, security measures, and overall resilience against common threats.

\textbf{Strengths:}
\begin{itemize}
\item Proposes a detailed forensic analysis of multiple apps.
\item Can guide developers in enhancing the security features of mobile banking apps.
\item Highlights specific vulnerabilities in the apps.
\end{itemize}

\textbf{Limitations:}
\begin{itemize}
\item Limit scope (App and geography).
\item Limited sample
\end{itemize}

\hrulefill
\newline

\noindent
\textbf{Paper Title \& Reference:} \textit{Security Issues of Unified Payments Interface and Challenges: Case Study}~\cite{9478078}

\textbf{Type of Study:} App analysis

\textbf{Summary:} The paper provides an in-depth analysis of the Unified Payments Interface (UPI), focusing on its operational framework, security vulnerabilities, and enhancements over time. It outlines how UPI facilitates mobile banking transactions through a user-friendly interface, utilizing Virtual Private Addresses (VPAs) and Payment \&Service Providers (PSPs). The paper discusses the architecture of UPI, comparing it with the Immediate Payment Service (IMPS), and highlights the security loopholes present in earlier versions, particularly BHIM UPI 1.0. It also addresses the improvements made in UPI 2.0 to mitigate these vulnerabilities and introduces the concept of offline UPI transactions.

\textbf{Strengths:}
\begin{itemize}
\item Provides a thorough explanation of how UPI works.
\item Highlights significant security issues and vulnerabilities in UPI, particularly in earlier versions.
\item Discusses UPI 2.0 and the measures taken to enhance security.
\item Discusses real-world implications of UPI's security issues.
\end{itemize}

\textbf{Limitations:}
\begin{itemize}
\item Lack of comprehensive security analysis for UPI 2.0.
\item Limit context and scope.
\item Lack of empirical evidence of the current security landscape for UPI.
\end{itemize}

\hrulefill
\newline

\noindent
\textbf{Paper Title \& Reference:} \textit{Mesh Messaging in Large-Scale Protests:
Breaking Bridgefy}~\cite{10.1007/978-3-030-75539-3_16}

\textbf{Type of Study:} App analysis

\textbf{Summary:} The paper provides a comprehensive analysis of the security vulnerabilities associated with Bridgefy, a mesh messaging application used in protest scenarios. In this work, authors reverse-engineered the Bridgefy platform using the Jadx tool, and they performed a dynamic inspection using the Frida toolkit to identify several critical weaknesses that undermine its security claims and its effectiveness in facilitating secure communication during protests. The paper details various avenues for tracking users and building social graphs, as well as the lack of effective authentication mechanisms, which could lead to impersonation and man-in-the-middle (MITM) attacks. Authors demonstrated how certain attacks, including variants of Bleichenbacher’s attack, can break confidentiality using chosen ciphertexts. They argued that the use of a "zip bomb" could disable the mesh network, highlighting the risks of relying on Bridgefy in critical situations.

\textbf{Strengths:}
\begin{itemize}
\item Provides a thorough examination of the Bridgefy application, including reverse engineering and vulnerability assessment.
\item The focus on the application’s use in protest scenarios makes the findings particularly relevant.
\item Outlines the responsible disclosure process.
\item Presents concrete examples of attacks that can be executed against the app.
\end{itemize}

\textbf{Limitations:}
\begin{itemize}
\item Limited analysis to a specific version of the app.
\item Not include empirical data on how users interact with the app.
\item Findings may not be generalizable to all mesh messaging apps.
\end{itemize}

\hrulefill
\newline

\noindent
\textbf{Paper Title \& Reference:} \textit{Security Analysis of Unified Payments Interface
and Payment Apps in India}~\cite{247668}

\textbf{Type of Study:} App analysis

\textbf{Summary:} In this paper, the authors present a comprehensive security analysis of the UPI protocol, which is widely used in India for digital payment apps. They focus on identifying vulnerabilities within the UPI system and the associated payment apps. The study involved reverse engineering the UPI protocol through various apps, as the protocol details were not publicly available. It uncovered significant weaknesses in the multi-factor authentication workflow of UPI 1.0, which could lead to severe security implications for users. Authors responsibly disclosed their findings and noted that subsequent updates to UPI 2.0 addressed some of the identified vulnerabilities, although several underlying issues remained.

\textbf{Strengths:}
\begin{itemize}
\item Provides a detailed security analysis of the UPI 1.0 protocol. 
\item Addresses a critical area of concern.
\item Employes a principled approach to reverse-engineer the UPI protocol.
\item The vulnerabilities identified are responsibly disclosed to encourage updates to UPI 2.0.
\end{itemize}

\textbf{Limitations:}
\begin{itemize}
\item Limited sample size.
\item The security defenses in the apps prevent the use of automated analysis techniques.
\item Limit scope (OS, app category, and geography).
\end{itemize}

\hrulefill
\newline

\noindent
\textbf{Paper Title \& Reference:} \textit{Forensic Analysis of Mobile Banking Applications in Nigeria}~\cite{Uduimoh}

\textbf{Type of Study:} App analysis

\textbf{Summary:} The paper focuses on the forensic examination of five Android-based mobile banking applications used in Nigeria. The primary objectives are to assess the amount of user data generated and retained by these applications after user registration and transactions and to determine if this data can be utilized to identify user actions or transactions. Findings reveal significant insights into the types of sensitive user data retained by these applications, which raises concerns about user privacy and data security.

\textbf{Strengths:}
\begin{itemize}
\item Addresses a gap in the literature regarding mobile banking apps in Nigeria.
\item Provides a broader understanding of the data retention practices across different platforms.
\item Findings are relevant for stakeholders.
\end{itemize}

\textbf{Limitations:}
\begin{itemize}
\item The sample size is not representative.
\item Limit scope (OS, app category, and geography).
\end{itemize}

\hrulefill
\newline

\noindent
\textbf{Paper Title \& Reference:} \textit{Mo(Bile) Money, Mo(Bile) Problems: Analysis
of Branchless Banking Applications}~\cite{Reaves}

\textbf{Type of Study:} App analysis

\textbf{Summary:} The paper evaluates the security of mobile money applications, particularly focusing on their vulnerabilities and the overall state of security in the mobile money ecosystem. The authors conducted a follow-up analysis nearly a year after their initial study to assess whether the security of these applications had improved. They found that the security of mobile money apps for Android had not significantly improved since 2015: many mobile money applications remained vulnerable to TLS man-in-the-middle attacks at a rate four times higher than normal apps. Nearly half of the apps contacted servers with insecure TLS configurations, and critical vulnerabilities identified in previous analyses had not been addressed, despite ongoing development of non-security features. The authors emphasize the need for collaboration among researchers, regulators, and developers to enhance the security of mobile money apps.

\textbf{Strengths:}
\begin{itemize}
\item Provides a thorough examination of mobile money applications.
\item Provides a longitudinal study.
\item Proposes collaboration with stakeholders.
\end{itemize}

\textbf{Limitations:}
\begin{itemize}
\item Limit scope (OS and app category).
\item Limitations in resources to reanalyze apps after vendors made changes.
\item Vendor response variability, which may impact the reliability of the findings.
\end{itemize}

\hrulefill
\newline

\noindent
\textbf{Paper Title \& Reference:} \textit{Side-Effects of Permissions Requested by Mobile Banking on Android
Platform: A Case Study of Morocco}~\cite{10.1145/3108421.3108433}

\textbf{Type of Study:} App analysis

\textbf{Summary:} The paper investigates the security issues about the permissions requested by mobile banking applications on the Android platform, focusing on Morocco as a case study. It highlights the potential dangers of these permissions, their effects on sensitive user data, and their relationship with the attack called “Man in the Middle” and its different forms. The authors analyze the permissions requested by the BaridBank Mobile application, comparing it with another banking application, Attijari Mobile, which requests no permissions. The paper also presents the results of an analysis of 100 mobile banking applications using a tool called “PerUpSecure”.

\textbf{Strengths:}
\begin{itemize}
\item Provides a comprehensive analysis of the permissions requested by mobile banking apps.
\item Uses a real-world case study (BaridBank Mobile), making the research relevant and applicable.
\item The use of the PerUpSecure tool to analyze a large set of applications adds credibility to the findings.
\end{itemize}

\textbf{Limitations:}
\begin{itemize}
\item Limit scope (OS, app category, and geography).
\item Analysis based on permissions requested, not how they are used or misused in practice.
\item Does not propose specific solutions or strategies to mitigate the identified risks.
\end{itemize}

\hrulefill
\newline

\noindent
\textbf{Paper Title \& Reference:} \textit{Vulnerability detection in recent Android apps: An empirical study}~\cite{7885802}

\textbf{Type of Study:} App analysis

\textbf{Summary:} The paper investigates the security vulnerabilities present in Android apps, particularly those developed by individual developers in Bangladesh. The authors conducted an empirical study by selecting a range of applications from a local app store and the Google Play Store and testing them with three different vulnerability detection tools. The study aimed to identify common vulnerabilities, understand their causes and suggest countermeasures to enhance app security. Findings reveal a high prevalence of vulnerabilities, particularly WebView vulnerabilities, which were found in 13 out of 19 tested apps, and other common vulnerabilities such as issues related to advertisement and storage access.

\textbf{Strengths:}
\begin{itemize}
\item Employs an empirical approach to analyze apps.
\item Selects apps from both local and popular sources.
\item Use of multiple tools to ensure a thorough examination of the apps.
\item Raises awareness regarding security practices.
\end{itemize}

Limitations:
\begin{itemize}
\item Focus on a small number of apps.
\item Limit scope (geographical focus).
\item Lack of in-depth analysis of the vulnerability implications.
\end{itemize}

\hrulefill
\newline

\noindent
\textbf{Paper Title \& Reference:} \textit{On the MitM vulnerability in mobile banking applications for android devices}~\cite{7947811}

\textbf{Type of Study:} App analysis

\textbf{Summary:} The paper discusses the security vulnerabilities in mobile banking applications, specifically focusing on the Man-in-the-Middle (MitM) attack. The authors tested 19 mobile banking applications deployed by public sector banks in India. They found that most of these apps were highly vulnerable to MitM attacks, even those using HTTPS for connection establishment. This indicates a poor implementation of the SSL framework in these applications. The paper also discusses associated attacks such as denial of service, session prediction, account lockout, and HTTP request smuggling.

\textbf{Strengths:}
\begin{itemize}
\item Provides a comprehensive analysis of the security vulnerabilities in mobile banking apps.
\item Covers a range of associated attacks for the identified vulnerabilities.
\item Provides suggestions about the requirements of addressing the basic security flaws.
\end{itemize}

\textbf{Limitations:}
\begin{itemize}
\item Limit scope (geography and app category).
\item Not provide specific recommendations or strategies for improving the app security.
\item The testing requires actual user credentials, which might raise ethical and privacy concerns.
\end{itemize}

\hrulefill
\newline

\noindent
\textbf{Paper Title \& Reference:} \textit{Signature Based Malicious Behavior Detection in
Android}~\cite{10.1007/978-981-15-6648-6_20}

\textbf{Type of Study:} App analysis and D\&I solutions

\textbf{Summary:} The paper presents a behavior-based approach for detecting Android malware by analyzing system-generated logs during the runtime of apps. The authors focus on identifying malicious behaviors that could lead to information leakage, privilege escalation, and unauthorized access to critical permissions. The proposed system aims to provide a more dynamic and insightful detection method compared to traditional static analysis techniques. It was tested on a variety of applications, including those flagged as malicious by the government of India.

\textbf{Strengths:}
\begin{itemize}
\item Proposes a runtime analysis for more accurate detection of malicious activities. 
\item The system generates signatures based on various malicious behaviors.
\item The approach is effective, according to its high accuracy rates. 
\item The detection system is scalable.
\end{itemize}

\textbf{Limitations:}
\begin{itemize}
\item Dependance on the system logs (which may vary across devices and versions).
\item Limited scope of testing (further testing on a broader dataset may be necessary).
\item The evasion techniques employed by sophisticated malware are not addressed.
\end{itemize}

\hrulefill
\newline

\noindent
\textbf{Paper Title \& Reference:} \textit{Integrating OAuth and Aadhaar with
e-Health care System}~\cite{9012487}

textbf{Type of Study:} D\&I solutions

\textbf{Summary:} The paper explores the integration of OAuth 2.0 (a secure protocol for access delegation) and Aadhaar (India’s unique identification system) authentication into e-Health apps in India to enhance security and streamline user authentication. The primary goal is to provide secure access to electronic health records (EHR) and facilitate efficient interaction between patients and healthcare providers. The paper outlines the implementation of the proposed system, including authentication, registration, and login flows. It also discusses offline access and the use of big data tools for analytics.

\textbf{Strengths:}
\begin{itemize}
\item Provides a robust authentication mechanism.
\item Aadhaar-based authentication simplifies the registration and login processes.
\item Covers various aspects of eHealth applications, from different types of apps to the roles of stakeholders.
\item The system is scalable with the use of cloud infrastructure and big data tools.
\end{itemize}

\textbf{Limitations:}
\begin{itemize}
\item Dependency on Aadhaar.
\item The implementation complexity may be a barrier for smaller healthcare providers.
\item Use of centralized storage.
\item The extent of the offline access functionality is not fully detailed.
\end{itemize}

\hrulefill
\newline

\noindent
\textbf{Paper Title \& Reference:} \textit{A
Comparative Study of User Data Security and Privacy in Native and
Cross Platform Android Mobile Banking Applications}~\cite{Ansong}

\textbf{Type of Study:} DFS

\textbf{Summary:} The study investigates the security and privacy of user data in Android mobile banking applications, based on permissions requested and used. The objective is to determine which development framework (native or cross-platform) ensures better user data privacy and security in mobile banking applications. The study focuses on 22 mobile banking apps from Ghanaian banks and identifies the development frameworks and permissions they requested/used. The findings reveal that cross-platform apps tend to use more of the permissions they request compared to native apps, which use fewer of the requested permissions. Consequently, cross-platform frameworks are recommended for mobile banking apps.

\textbf{Strengths:}
\begin{itemize}
\item Provides a detailed comparison of permissions requested and used by both native and cross-platform apps.
\item The use of specific tools and a structured approach adds credibility to the findings.
\item Offers recommendations for developers and banks to enhance security and privacy in mobile banking apps.
\end{itemize}

\textbf{Limitations:}
\begin{itemize}
\item The AVC UnDroid tool could not analyze apps larger than 14MB.
\item The sample size may be not representative.
\item Limit scope (geography and app category).
\item Focus only on Apache Cordova for cross-platform apps.
\end{itemize}